\documentstyle[12pt,epsf]{article}
\begin{document}
\addtolength{\baselineskip}{.7mm}
\thispagestyle{empty}
\begin{flushright}
TIT/HEP--290 \\
{\tt hep-th/9504150} \\
April, 1995
\end{flushright}
\vspace{2mm}
\begin{center}
{\Large{\bf Mass Spectra of Supersymmetric Yang-Mills Theories
in $1 + 1$ Dimensions}} \\[15mm]
{\sc Yoichiro Matsumura}\footnote{
present address: Central Training Center, Meidensha Corporation, \\
\indent \ Numazu, Shizuoka 410 Japan},
{\sc Norisuke Sakai}\footnote{
\tt e-mail: nsakai@th.phys.titech.ac.jp}  \\[3mm]
and \\[3mm]
{\sc Tadakatsu Sakai}\footnote{
\tt e-mail: tsakai@th.phys.titech.ac.jp} \\[4mm]
{\it Department of Physics, Tokyo Institute of Technology \\[2mm]
Oh-okayama, Meguro, Tokyo 152, Japan}  \\[15mm]
{\bf Abstract}\\[5mm]
{\parbox{13cm}{\hspace{5mm}
Physical mass spectra of supersymmetric Yang-Mills theories in
$1+1$ dimensions are evaluated in the light-cone gauge with a
compact spatial dimension.
The supercharges are constructed and
the infrared regularization is unambiguously prescribed for
supercharges, instead of the light-cone Hamiltonian.
This provides a manifestly supersymmetric infrared regularization
for the discretized light-cone approach.
%The mass spectra is obtained b
By an exact
diagonalization of the supercharge matrix between up to
several hundred color singlet bound states, we find a rapidly increasing
density of states as mass increases.
}}
\end{center}
\vfill
\newpage
\setcounter{section}{0}
\setcounter{equation}{0}
\setcounter{footnote}{0}
%\numberbysection
%\def\theequation{\arabic{section}.\arabic{equation}}
%
%
%%%%%%%  Section 1  %%%%%%%%%%%%%%%%%%%%%%%%%%%%%%%%%%%%%%%%
%
\vspace{7mm}
\pagebreak[3]
\addtocounter{section}{1}
\setcounter{equation}{0}
\setcounter{subsection}{0}
\setcounter{footnote}{0}
\begin{center}
{\large {\bf \thesection. Introduction}}
\end{center}
\nopagebreak
\medskip
\nopagebreak
\hspace{3mm}

Supersymmetric theories offer promising models for the
unified theory.
Both as a model for grand unified theories and as a low energy
effective theory for superstring, the dynamics of supersymmetric
Yang-Mills %(YM)
gauge theories is a fascinating subject.
The nonperturbative aspects of supersymmetric theories are
 crucial to understand fundamental aspects of such
theories, especially the supersymmetry breaking.

One of the most popular models for the supersymmetry breaking
is currently to assume the gaugino bilinear condensation in
the supersymmetric Yang-Mills gauge theories  \cite{Nilles}.
Although the condensation itself may not break supersymmetry
in the supersymmetric gauge theories, it will give rise to
the supersymmetry breaking if embedded in supergravity \cite{VeYa}.
Since the fermion bilinear condensation is implied by
 the chiral symmetry breaking in QCD,
one can expect a similar
nonperturbative effects in supersymmetric Yang-Mills gauge theories.
Moreover, recent progress in understanding duality in supersymmetric
Yang-Mills gauge theories opened up a rich arena for studying the
nonperturbative effects in supersymmetric gauge
theories \cite{SeWi}.

It has been quite fruitful to study Yang-Mills gauge theories in $1+1$
dimensions
instead of studying directly the four dimensional counterpart.
In $1+1$ dimensions, Yang-Mills gauge field itself has no dynamical
degree of
freedom as a field theory, but gives rise to a confining potential
for colored particles \cite{THooft}.
Many aspects of color singlet bound states can be explored by solving
the theory in the large $N$ limit \cite{CaCoGr}.
Unfortunately the supersymmetric gauge multiplet contains genuine
dynamical degree of freedom in the adjoint representation of the
gauge group contrary to ordinary Yang-Mills gauge theory \cite{Ferrara}.
Therefore one cannot obtain a simple closed form for the color singlet
bound states even in the large $N$ limit.

There has been progress in studying the dynamics of matter fields in the
adjoint representation in ordinary Yang-Mills gauge
theories \cite{DaKl}.
They have used the light-cone quantization and compactified the
spatial dimension to give discrete momenta.
In this discretized light-cone quantization approach,
one can diagonalize the mass matrix for finite number of light-cone
momenta and can hope to obtain the infinite volume limit
eventually \cite{PuBr}, \cite{HoBr}.
The Yang-Mills gauge theory with only the adjoint matter fermion
is used to propose a kind of supersymmery which
is valid only at a particular value of a parameter and
is different from the usual
linearly realized supersymmetry \cite{Kutasov}.
More recently, gauge theories in $1+1$ dimensions
with matter in adjoint representations
was studied focusing attention on zero modes \cite{LeShTh}.
The zero modes are generally important in revealing nontrivial
vacuum structures such as the vacuum condensate \cite{MaYa}.

In spite of these investigations of Yang-Mills gauge theories with
adjoint scalar and spinor matter fields, there are two points
which necessitate a new analysis of physical spectra
in the case of supersymmetric
gauge theories.
The first point is that the coexistence of spinor and scalar
gives rise to a large number of new ``mixed'' physical states,
partly consisting of spinors and partly of scalars as
constituents.
The second point is the presence of a specific amount of
the Yukawa interaction
which is a distinguishing feature of the supersymmetric Yang-Mills
theory \cite{Ferrara}.

The purpose of our paper is to study the supersymmetric Yang-Mills
gauge theories
in $1+1$ dimensions through the discretized light-cone quantization.
We construct the supercharge explicitly and specify an
infrared regularization for supercharge by means of
the discretized version of the principal value prescription.
By using the supercharge, we succeed in overcoming ambiguities
in prescribing the infrared regularization for the light-cone
Hamiltonian.
As a result, the regularization preserves the supersymmetry algebra
manifestly.
For light-cone momenta
up to 8 units of the smallest momentum,
we find several hundred color singlet bound states of bosons and the
same number of fermions.
We exactly diagonalize the supercharge instead of the Hamiltonian to
obtain masses, degeneracies, and the average number of constituents
in these bound states.
We observe that the density of the bound states as a function of their
masses tends to converge in the large volume limit.
It is consistent with the rapidly increasing density of states
suggested by the closed string interpretation.
Since we preserve supersymmetry at each stage of
our study, we naturally obtain exact correspondence
between bosonic and fermionic color singlet bound states.
Although we postpone studing the issue of zero modes,
our results in the present approximation suggest that supersymmetry
is not broken in this supersymmetric Yang-Mills gauge theory.
It is an interesting future problem to see if our supersymmetric
theory can offer a model for gaugino condensation.
For that purpose, one should study the zero mode
in this theory.
However, the present investigation is focused on physical mass spectra
as a first step to understand the dynamics of the supersymmetric
Yang-Mills theories.

Before writing up our paper, we have received a
preprint, where the same theory has been studied
by means of the Makeenko-Migdal loop equations \cite{Li}.
With certain assumptions, the author gave an interesting solution
and also argued for the nonvanishing Witten index.
Although his method is worth exploring, it may not be suitable
to obtain physical quantities such as mass spectra.
In this respect, our methods are complimentary to his, and
our conclusions are consistent with each other.

In sect. 2, supersymmetric Yang-Mills gauge theories in $1+1$
dimensions are quantized
and supercharges are defined.
In sect. 3, the compact spatial
dimension is introduced in the light-cone quantization and the
supercharges are discretized.
The result of our exact diagonalization of supercharge is presented
and discussed in sect. 4.
Superfields and supertransformations
 are summarized in Appendix A.
Truncation of the bound state equation to the two constituents
subspace is given in Appendix B.
Explicit mass matrices with mass terms for adjoint scalar and spinor
is given in Appendix C.
%
%%%%%%%  Section 2  %%%%%%%%%%%%%%%%%%%%%%%%%%%%%%%%%%%%%%%%
%
\vspace{7mm}
\pagebreak[3]
\addtocounter{section}{1}
\setcounter{equation}{0}
\setcounter{subsection}{0}
\setcounter{footnote}{0}
\begin{center}
{\large {\bf \thesection. SUSY Yang-Mills Theories in $1+1$ Dimensions}}
\end{center}
\nopagebreak
\medskip
\nopagebreak
\hspace{3mm}

In two-dimensions, the gauge field $A^\mu$ is contained in a
supersymmetric multiplet consisting of a Majorana fermion $\Psi$
and a scalar $\phi$ in the adjoint representation of the gauge
group together with gauge field itself \cite{Ferrara}.
Therefore our field content is different from that in ref \cite{Kutasov}.
%We summarize the superfield formalism in Appendix A.
After choosing the Wess-Zumino gauge, we have an
action
\begin{equation}
S =
\int d^2 x \, {\rm tr} \Bigg[-{1\over 4 g^{2}} F_{\mu \nu}F^{\mu \nu}
+\frac{1}{2} D_{\mu}\phi D^{\mu}\phi
+i\bar{\Psi} \gamma^{\mu} D_{\mu} \Psi \nonumber
  -2ig\phi\bar{\Psi} \gamma_{5} \Psi \Bigg],
\label{eq:sacda}
\end{equation}
where
$A_{\mu}$,
$\phi$,
$\Psi$,
and
$\bar \Psi=\Psi^T \gamma^0$
are traceless $N\times N$ hermitian matrix for $U(N)$ ($SU(N)$) gauge
group,
 $g$ is the gauge coupling constant,
$F_{\mu\nu} = \partial_{\mu}A_{\nu} - \partial_{\nu}A_{\mu}+i[A_{\mu}
A_{\nu}]$ and $D_{\mu}$ is the usual covariant derivative
\begin{equation}
D_{\mu}\phi = \partial_{\mu}\phi+i[A_{\mu}, \phi], \quad
D_{\mu}\Psi = \partial_{\mu}\Psi+i[A_{\mu}, \Psi] .
\end{equation}
The supersymmetry dictates the presence of the Yukawa type interaction
between the adjoint spinor and scalar fields with the strength of
the gauge coupling.
The supersymmetric Yang-Mills gauge theory in two-dimensions
can be obtained by a dimensional reduction from the supersymmetric
Yang-Mills gauge theory in three dimensions.
The adjoint scalar field can be understood as the component of the gauge
field in the compactified dimension and the Yukawa coupling is
nothing but the gauge interaction in this compactified extra dimension.

In the Wess-Zumino gauge, the remaining invariances
of the action are the usual gauge invariance and a supertransformation
which is obtained by combining the supertransformation
and the compensating gauge transformation in the superfield
formalism as summarized in the Appendix A.
We denote this modified supertransformation as $\tilde{\delta}_{super}$
which is given in terms of component fields as
($\epsilon^{01}=-\epsilon_{01}=1$)
\begin{eqnarray}
\tilde{\delta}_{super}A_{\mu}
&\!\!\!
=
&\!\!\!
ig\bar{\epsilon}\gamma_{5}
\gamma_{\mu}\sqrt{2}\Psi , \quad
\tilde{\delta}_{super}\phi = -\bar{\epsilon}\sqrt{2}\Psi , \nonumber \\
\tilde{\delta}_{super} \Psi
&\!\!\!
=
&\!\!\!
 \displaystyle -\frac{1}{2\sqrt{2}g}
\epsilon \epsilon^{\mu\nu}F_{\mu\nu}
+ \frac{i}{\sqrt{2}}\gamma^{\mu}\epsilon D_{\mu}\phi .
\label{eq:st}
\end{eqnarray}
The corresponding spinor supercurrent $j^\mu$ is given by
\begin{equation}
\bar \epsilon j^\mu =
\, {\rm tr}\left[-\sqrt{2}\bar \epsilon \Psi D^\mu \phi+i{1 \over \sqrt{2}g}
 \epsilon^{\nu\lambda}F_{\nu\lambda}
 \bar \epsilon \gamma^\mu \Psi+\sqrt{2} \bar \epsilon \gamma_5
\Psi \epsilon^{\mu\nu} D_\nu \phi\right] .
\label{supercurrent}
\end{equation}

We introduce the light-cone coordinates where the line
element $ds^2$ is given by
\begin{equation}
x^{\pm}=\frac{1}{\sqrt{2}}(x^{0}\pm x^{1}),
\quad
ds^2=(dx^0)^2-(dx^1)^2=2dx^+dx^-.
\end{equation}
We decompose the spinor and use gamma matrices
\begin{equation}
\Psi_{ij}=2^{-1/4} (\psi_{ij},\chi_{ij})^{T},
\qquad
\gamma^0 =\sigma_2,\; \gamma^1 =i\sigma_1,\;
\gamma_5 = \gamma^0 \gamma^1=\sigma_3.
\label{gammamatrix}
\end{equation}
Taking the light-cone gauge
$
A_{-} = A^{+} =0
$ and $x^+$ as time, we find the action
\begin{eqnarray}
S &\!\!\!=&\!\!\!
\int dx^{+}dx^{-} \, {\rm tr} \Bigg[\partial_{+}\phi\partial_{-}\phi
+i\psi\partial_{+}\psi +i\chi\partial_{-}\chi
\nonumber \\
&\!\!\!&\!\!\!
+\frac{1}{2g^2} (\partial_{-}A_{+})^2
+A_{+}J^{+}
+\sqrt{2}g\phi\{\psi,\chi\}
 \Bigg],
\label{eq:action}
\end{eqnarray}
where the current $J^+$
receives
contributions from the scalar $J_{\phi}^{+}$ and the spinor
$J_{\psi}^{+}$
\begin{equation}
J^+=J_{\phi}^{+} +
J_{\psi}^{+}, \quad
J_{\phi}^{+} = i[\phi,\partial_{-}\phi], \quad
J_{\psi}^{+} = 2\psi\psi.
\label{totalcurrent}
\end{equation}

We do not need Faddeev-Popov ghosts
in this gauge.
Since the action contains no time derivative for the
gauge potential $A_{+}$ and the left-moving fermion $\chi$,
they can be eliminated by means of constraints obtained as
their Euler-Lagrange equations
\begin{equation}
i\sqrt{2} \partial_{-}\chi
- g[\phi,\psi] = 0,
\quad
\partial_{-}^{2} \bar{A}_{+} - g^2 J^{+} = 0.
\label{eq:const}
\end{equation}
where $\bar{A}_{+}$ is the non-zero mode of $A_{+}$.
The zero mode of $A_{+}$ plays
the role of a Lagrange multiplier which provides a constraint
\begin{equation}
\int dx^- J^+ =0.
\label{shiki2}
\end{equation}
This constraint will give a restriction for physical states in quantum
theory.
After eliminating the fields $A_+$ and $\chi$,
we find that the action becomes
\begin{equation}
S = \int dx^{+}dx^{-} \, {\rm tr} \Bigg[  \partial_{+}\phi\partial_{-}\phi
+i\psi\partial_{+}\psi
 +\frac{g^2}{2}J^{+}\frac{1}{\partial_{-}^{2}}J^{+}
-\frac{1}{2}ig^2 [\phi,\psi]\frac{1}{\partial_{-}} [\phi,\psi] \Bigg].
\label{eq:actionphys}
\end{equation}
Let us note that the constraints give rise to non-local terms
in the action.

By the Noether procedure, we construct the energy momentum
tensor $T^{\mu\nu}$, and light-cone momentum and energy
$P^{\pm}= \int dx^{-} T^{+\pm}$ on a constant light-cone time
\begin{equation}
P^{+} =  \int dx^{-}
\, {\rm tr}\biggl[(\partial_{-}\phi)^{2} +i\psi\partial_{-}\psi \biggl],
\label{eq:p+}
\end{equation}
\begin{equation}
P^{-} =  \int dx^{-}
\, {\rm tr} \Bigg[- \frac{g^2}{2}  J^{+} \frac{1}{\partial_{-}^{2}} J^{+}
+\frac{i}{2} g^2 [\phi,\psi]\frac{1}{\partial_{-}}[\phi,\psi] \Bigg].
\label{eq:p-}
\end{equation}

The supercharges $Q_1$ and $Q_2$
are defined as integrals of the upper and
lower components of the spinor supercurrent $j^\mu=(j_1^\mu, j_2^\mu)$
in eq.(\ref{supercurrent})
\begin{equation}
Q_1 \equiv \int dx^- j_1^+
=
 2^{1/4}\int dx^{-}\, {\rm tr}\left[\phi\partial_{-}\psi-\psi\partial_{-}\phi
\right] ,
\end{equation}
\begin{eqnarray}
Q_2
&\!\!\!
\equiv
&\!\!\!
\int dx^- j_2^+
=
 2^{3/4}g\int dx^{-}\, {\rm tr}\left[J^+{1 \over \partial_-}\psi\right]
\nonumber \\[.5mm]
&\!\!\!
=
&\!\!\!
 2^{3/4}g\int dx^{-}\, {\rm tr}
\left\{\left(i[\phi, \partial_-\phi] + 2\psi\psi\right)
{1 \over \partial_-}\psi\right\} .
\end{eqnarray}

Using the conjugate momenta
$\pi_{\phi}= \partial {\cal L}/\partial (\partial_{+}\phi)
= \partial_{-}\phi$
for adjoint scalar field $\phi_{ij}$
and $\pi_{\psi}= \partial {\cal L}/\partial (\partial_{+}\psi)
= i\psi$
for adjoint spinor field $\psi_{ij}$,
the canonical (anti)commutation relation
are given at equal light-cone time $x^{+} = y^{+}$ by
\begin{equation}
[\phi_{ij}(x),\partial_{-}\phi_{kl}(y)]
=
i \{\psi_{ij}(x), \psi_{kl}(y)\}
=
\frac{1}{2}i \delta(x^{-}-y^{-})\delta_{il} \delta_{jk}.
\label{cancomrel}
\end{equation}

We expand the fields
in modes with momentum $k^+$
at light-cone time $x^+=0$
\begin{equation}
\phi_{ij}(x^{-} ,0) =  \frac{1}{\sqrt{2\pi}} \int_{0}^{\infty}
\frac{dk^{+}}{\sqrt{2k^{+}}}
\left( a_{ij}(k^{+}){\rm e}^{-ik^{+}x^{-}} +
a_{ji}^{\dag}(k^{+}) {\rm e}^{ik^{+}x^{-}}\right) ,
\label{eq:modephi}
\end{equation}
\begin{equation}
\psi_{ij}(x^{-} ,0) =  \frac{1}{2\sqrt{\pi}} \int_{0}^{\infty} dk^{+}
\left (b_{ij}(k^{+}){\rm e}^{-ik^{+}x^{-}} +
b_{ji}^{\dag}(k^{+}){\rm e}^{ik^{+}x^{-}}\right ).
\label{eq:modepsi}
\end{equation}
The canonical (anti-)commutation relations (\ref{cancomrel})
are satisfied by
\begin{equation}
[a_{ij}(k^{+}),a_{lk}^{\dag}(\tilde{k}^{+})]   =
\{b_{ij}(k^{+}), b_{lk}^{\dag}(\tilde{k}^{+})\}   =
\delta(k^{+} - \tilde{k}^{+}) \delta_{il} \delta_{jk}.
\end{equation}

In nonsupersymmetric theories, one can define finite
Hamiltonian operators only after discarding the usually divergent
vacuum energies \cite{DaKl}.
However, we should not discard any vacuum energies in supersymmetric
theories, since vacuum energies have an absolute meaning in
supersymmetric theories as an indicator of supersymmetry breaking.
In fact we will not need to discard the vacuum energies by hand,
provided we exercise care with respect to ordering of
operators.

One can obtain the light-cone momentum $P^{+}$ in terms of oscillators
\begin{equation}
P^{+} = \int_{0}^{\infty} dk k \left\{ a_{ij}^{\dag}(k)a_{ij}(k)
+b_{ij}^{\dag}(k)b_{ij}(k) \right\},
\end{equation}
where we dropped the superscript + on $k^{+}$ for brevity,
and henceforth we do so.

The light-cone Hamiltonian $P^{-}$ can be divided into two parts:
the current-current interaction term $P_{JJ}^{-}$ and
the Yukawa coupling term $P_{Yukawa}^{-}$
\begin{equation}
P^{-} = P_{JJ}^{-} + P_{Yukawa}^{-} .
\label{shiki3}
\end{equation}
Let us introduce the momentum representation of
the current $J^{+}$
\begin{eqnarray}
\tilde{J}^{+}(k) = \frac{1}{\sqrt{2\pi}}\int dx^{-} J^{+}(x^{-})
\exp(-ikx^{-}) .
\label{eq:jda}
\end{eqnarray}
Substituting the mode expansions (\ref{eq:modephi}) and
(\ref{eq:modepsi}), we obtain for $-k < 0$
\begin{eqnarray}
&\!\!\!
&\!\!\!
\tilde{J}^{+}_{ij}(-k)
=
 \frac{1}{2\sqrt{2\pi}}\int_{0}^{\infty} dp
\frac{2p + k}{\sqrt{p(p+k)}}\left[a^{\dag}_{ki}(p)a_{kj}(k+p)
-a^{\dag}_{jk}(p)a_{ik}(k+p)\right] \nonumber \\
&\!\!\!
+
&\!\!\!
 \frac{1}{2\sqrt{2\pi}}\int_{0}^{k} dp \frac{k-2p}{\sqrt{p(k-p)}}
a_{ik}(p)a_{kj}(k-p)  \\
&\!\!\!
+
&\!\!\!
\frac{1}{\sqrt{2\pi}}\int_{0}^{\infty} dp
\left[b^{\dag}_{ki}(p)b_{kj}(k+p)
-b^{\dag}_{jk}(p)b_{ik}(k+p)\right]
+ \frac{1}{\sqrt{2\pi}}\int_{0}^{k} dp
b_{ik}(p)b_{kj}(k-p) . \nonumber
\label{eq:cur}
\end{eqnarray}
Note that $\tilde{J}_{ij}^{+}(k) =
\left[\tilde{J}_{ji}^{+}(-k)\right]^{\dag}$ .
There are no (divergent) c-number terms in $\tilde{J}^{+}(k)$
since supersymmetry requires for the bosonic and fermionic c-number
contributions to cancel each other.
Therefore $\tilde{J}^{+}(k)$ is just the same as the normal ordered
product $:\tilde{J}^{+}(k):$ .
The current-current interaction term is given by
\begin{equation}
P^{-}_{JJ}
= \frac{g^{2}}{2}\int_{-\infty}^{\infty}\frac{dk}{k^{2}}
\tilde{J}^{+}_{ij}(k)\tilde{J}^{+}_{ji}(-k)
= \frac{g^{2}}{2}\int_{0}^{\infty}\frac{dk}{k^{2}}
 \left\{\tilde{J}^{+}_{ij}(k) , \tilde{J}^{+}_{ji}(-k) \right\} .
\label{eq:pphiphi}
\end{equation}

The source for $\chi$ is given
in the momentum space for $-k < 0$
\begin{eqnarray}
\widetilde{\left[\phi , \psi\right]}_{ij}(-k)
&\!\!\!
 =
&\!\!\!
\frac{1}{2\sqrt{2\pi}}\int_{0}^{\infty} dp \Bigg\{
\frac{1}{p}\left[a^{\dag}_{ki}(p)b_{kj}(k+p)
-a^{\dag}_{jk}(p)b_{ik}(k+p)\right] \nonumber \\
&\!\!\!
 +
&\!\!\!
\frac{1}{k+p}\left[a_{ik}(k+p)b^{\dag}_{jk}(p)
-a_{kj}(k+p)b^{\dag}_{ki}(p)\right] \Bigg\} \\
&\!\!\!
+
&\!\!\!
 \frac{1}{2\sqrt{2\pi}}\int_{0}^{k} dp \frac{1}{\sqrt{p}}
\left[a_{ik}(p)b_{kj}(k-p) - a_{kj}(p)b_{ik}(k-p) \right] ,\nonumber
\label{eq:phipsi}
\end{eqnarray}
Note that $\widetilde{\left[\phi , \psi\right]}_{ij}(k) =
-\left[\widetilde{\left[\phi , \psi\right]}_{ji}(-k) \right]^{\dag}$.
The Yukawa coupling term is given by
\begin{equation}
P^{-}_{Yukawa}
= -\frac{g^{2}}{2} \int_{0}^{\infty} \frac{dk}{k}\left[
\widetilde{[\phi,\psi]}_{ij}(k) \:,\:
\widetilde{[\phi,\psi]}_{ji}(-k)\right].
\end{equation}
When we bring the Hamiltonian into a normal ordered form, we find
that the (divergent) c-number vacuum energies cancel between bosons
and fermions.
Moreover, the only additional term $P_{quad}^-$ compared to normal
ordered Hamiltonian $:P^-:$ is quadratic
and is symmetric between
scalar and spinor oscillators
\begin{eqnarray}
&\!\!\!
&\!\!\!
P^-
=
P_{quad}^- + :P^-: \\
P_{quad}^-
&\!\!\!
=
&\!\!\!
{g^2 \over 4\pi}\int_0^{\infty} {dk \over k} C(k)
\left(a_{ij}^\dagger(k) a_{lm}(k)+b_{ij}^\dagger(k) b_{lm}(k)\right)
(N\delta_{il} \delta_{jm} -\delta_{ij}\delta_{lm})
\end{eqnarray}
\begin{equation}
C(k) \equiv \int_0^k dp \left({4k \over p^2}+{1 \over p}\right)
=\int_0^k dp {(k+p)^2 \over p(k-p)^2}
\label{quadcoefficient}
\end{equation}

The supercharge is given in terms of these operators as
\begin{equation}
Q_1 =
i 2^{1/4}
\int_0^{\infty} dk\sqrt{k}\left[a_{ij}(k)b_{ij}^{\dag}(k)
-a_{ij}^{\dag}(k)b_{ij}(k)\right] ,
\label{superchargeone}
\end{equation}
\begin{equation}
Q_2 =
-i 2^{1/4}
g\int_0^{\infty} {dk \over k}
\left[b_{ij}^{\dagger}(k) \tilde J_{ij}(-k)
-\left(\tilde J_{ij}(-k)\right)^{\dagger}b_{ij}(k)\right] .
\label{superchargetwo}
\end{equation}
By taking (anti-)commutators with spinor $\psi$ and scalar $\phi$
fields, we can confirm that these supercharge operators generate
supertransformations in the light-cone gauge as given in
(\ref{lightconesuperphi}) and (\ref{lightconesuperpsi}).

Our next task is to determine physical states whose mass spectra will be
calculated later. The light-cone vacuum is the Fock vacuum defined by
\begin{equation}
a_{ij}(k)\left|0\right\rangle = 0,
\quad
b_{ij}(k)\left|0\right\rangle = 0.
\end{equation}
satisfying $P^{\pm}\left|0\right\rangle = 0$ .
Fock states are given by acting creation operators $a_{ij}^\dagger (k) ,
b_{ij}^\dagger (k)$ and their linear combinations on
$\left|0\right\rangle$ .
In leading order in the $1/N$ expansion, physical states are given by
gauge singlet states with single trace of creation operators
$
\, {\rm tr}\left[{\cal O}(k_{1})\cdots {\cal O}(k_{m})\right]
\left|0\right\rangle / N^{m/2}\sqrt{s}
$
with ${\cal O}(k)$ representing $a^{\dag}(k)$ or $b^{\dag}(k)$, and
$N^{-m/2}$ the normalization factor and $s$ a symmetry factor.

The mass spectrum is obtained by solving the
eigenvalue problem
\begin{equation}
2P^{+}P^{-}\left|\Phi\right\rangle = M^{2}\left|\Phi\right\rangle .
\label{eq:eigen}
\end{equation}
It is almost impossible to solve the eigenvalue
problem analytically because we must diagonalize an
infinite dimensional matrix.
Therefore we will resort to a discretized approximation in the
next section.
Truncation to two constituent subspace yields a closed bound state
equation similar to the 'tHooft equation \cite{THooft} as described
in Appendix B.

%
%%%%%%%  Section 3  %%%%%%%%%%%%%%%%%%%%%%%%%%%%%%%%%%%%%%%%
%
\vspace{7mm}
\pagebreak[3]
\addtocounter{section}{1}
\setcounter{equation}{0}
\setcounter{subsection}{0}
\setcounter{footnote}{0}
\begin{center}
{\large
{\bf \thesection. Discretized Light-Cone Quantization of Superchage}}
\end{center}
\nopagebreak
\medskip
\nopagebreak
\hspace{3mm}

In order to prescribe the infrared regularization precisely and
to evaluate the mass spectrum in spaces with finite number of physical
states, we compactify spatial direction $x^{-}$
to form a circle with radius $2L$
by identifying $x^-=0$ and $x^-=2L$.
In order to preserve supersymmetry, we need to impose the same
boundary condition on scalars $\phi_{ij}$ and spinors $\psi_{ij}$.
It is in general necessary to choose periodic boundary conditions
on bosonic field and to retain zero modes,
if one wishes to take into account the possibility of
 vacuum condensate or spontaneous symmetry breaking \cite{MaYa}.
Since we are primarily interested in physical mass spectrum,
we neglect the zero modes in the present work.
We shall choose periodic boundary
conditions for both scalars $\phi_{ij}$ and spinors $\psi_{ij}$,
leaving the problem of zero modes for a further study
\begin{equation}
\phi_{ij}(x^{-}) = \phi_{ij}(x^{-}+2L),\quad
\psi_{ij}(x^{-}) = \psi_{ij}(x^{-}+2L).
\end{equation}
The allowed momenta become discrete and the momentum integral
is replaced by a summation,
\begin{equation}
k_{n}^{+} = \frac{\pi}{L} n ,\quad n=1,2,3,....\:, \qquad
\int_{0}^{\infty} dk^{+} \rightarrow
 \frac{\pi}{L}\sum_{n =1}^{\infty}.
\label{momentumsum}
\end{equation}
Then mode expansions (\ref{eq:modephi}) and (\ref{eq:modepsi})
for $\phi_{ij}$ and $\psi_{ij}$ become discretized
\begin{equation}
\phi_{ij} =  \frac{1}{\sqrt{4\pi}} \sum_{n=1}^{\infty}
\frac{1}{\sqrt{n}} \bigg[ A_{ij}(n){\rm e}^{-i\pi n x^{-}/L} +
A_{ji}^{\dag}(n) {\rm e}^{i\pi nx^{-}/L}\bigg],
\end{equation}
\begin{equation}
\psi_{ij} =  \frac{1}{\sqrt{4L}}
\sum_{n=1}^{\infty} \bigg[ B_{ij}(n){\rm e}^{-i\pi nx^{-}/L} +
B_{ji}^{\dag}(n) {\rm e}^{i\pi nx^{-}/L}\bigg],
\end{equation}
\begin{equation}
A_{ij}(n) = \sqrt{\pi /L} a_{ij}(k^{+}=\pi n/L),\quad
B_{ij}(n) = \sqrt{\pi /L} b_{ij}(k^{+}=\pi n/L),
\label{discreteoscillator}
\end{equation}
\begin{equation}
\left[A_{ij}(n),A_{lk}^{\dag}(n')\right] =
\left\{B_{ij}(n),B_{lk}^{\dag}(n')\right\} =
\delta_{nn'}\delta_{il}\delta_{jk} .
\end{equation}

Let us define the supercharge in this discretized light-cone
quantization.
The first supercharge $Q_1$ in
eq.(\ref{superchargeone})
in this compactified space is given by
\begin{equation}
Q_1 =
 2^{1/4}i \sqrt{\pi \over L}\sum_{n=1}^{\infty} \sqrt{n}
\left[A_{ij}(n)B_{ij}^{\dag}(n)
-A_{ij}^{\dag}(n)B_{ij}(n)\right] .
\label{discretesuperchargeone}
\end{equation}
Since the elimination of gauge field $A_+$ introduces a singular
factor $1/\partial_-$ in supercharge $Q_2$ in eq.(\ref{superchargetwo}),
we need to specify
an infrared regularization for this factor.
Following the procedure of 'tHooft \cite{THooft}, we
employ the principal value prescription for the supercharge.
Namely we simply drop the zero momentum mode
\begin{eqnarray}
Q_2 &\!\!\!
=&\!\!\!
 2^{1/4}(-i)g\sqrt{L \over \pi}\sum_{m=1}^{\infty} \frac{1}{m}
\left[B_{ij}^{\dagger}(m) \tilde J_{ij}(-m)
-\left(\tilde J_{ij}(-m)\right)^{\dagger}B_{ij}(m)\right]\nonumber \\
&\!\!\!
=
&\!\!\!
-i\frac{2^{-1/4}g}{\pi}\sqrt{L}\Bigg(
\sum_{l,n=1}^{\infty}\frac{l+2n}{2l\sqrt{n(l+n)}}  \biggl[
\left(A^{\dagger}(n)B^{\dagger}(l)
-B^{\dagger}(l)A^{\dagger}(n)\right)_{ij}A_{ij}(l+n) \nonumber \\
&\!\!\!
&\!\!\!
\;\;\;\;\;\;\;\;\;\;\;\;\;\;\;\;\;\;
-A_{ij}^{\dagger}(l+n)\Bigl(A(n)B(l)
-B(l)A(n)\Bigr)_{ij}\biggr]\nonumber \\
&\!\!\!
+
&\!\!\!
\sum_{l=3}^{\infty}\sum_{n=1}^{l-1}\frac{l-2n}{2l\sqrt{n(l-n)}}
\left[B_{ij}^{\dagger}(l)\biggl(A(n)A(l-n)\biggr)_{ij}
-\Bigl(A^{\dagger}(n)A^{\dagger}(l-n)\Bigr)_{ij}B_{ij}(l)
\right]\nonumber \\
&\!\!\!
-
&\!\!\!
\sum_{l,n=1}^{\infty}\left(\frac{1}{l}
+\frac{1}{n}\right)\biggl[\left(
B^{\dagger}(n)B^{\dagger}(l)\right)_{ij}B_{ij}(l+n)
+B_{ij}^{\dagger}(l+n)\Bigl(B(n)B(l)\Bigr)_{ij}\biggr]\nonumber \\
&\!\!\!
+
&\!\!\!
\sum_{l=2}^{\infty}\sum_{n=1}^{l-1}\frac{1}{l}\left[
B_{ij}^{\dagger}(l)\Bigl(B(n)B(l-n)\Bigr)_{ij}
+\left(B^{\dagger}(n)B^{\dagger}(l-n)\right)_{ij}B_{ij}(l)\right]\Bigg).
\label{discretesuperchargetwo}
\end{eqnarray}

The supersymmetry algebra requires a relation between
supercharges and the light-cone momentum $P^+$
and the Hamiltonian $P^-$ operators
\begin{equation}
\{Q_1, Q_1\}=2\sqrt2 P^+ ,
\label{superalgebraone}
\end{equation}
\begin{equation}
\{Q_2, Q_2\}=2\sqrt2 P^- ,
\label{superalgebratwo}
\end{equation}
\begin{equation}
\{Q_1, Q_2\}=0 ,
\label{superalgebrathree}
\end{equation}
in our choice of spinor notations (\ref{gammamatrix}).
Infrared regularizations of $P^+$ and $P^-$ have to be done
consistently with the supersymmetry algebra.
It is actually difficult to guess the correct infrared regularization
for the Hamiltonian unless we start from the supercharge.
The Hamiltonian $P^-$ can be defined by just squaring the supercharge
$Q_2$.
Then the above principal value prescription for the supercharge
 $Q_2$ specifies uniquely the prescription for the Hamiltonian.
In this way we can check that the supersymmetry algebra holds
in our formulation of the discretized light-cone quantization.

Physical states take the following form
\begin{equation}
\frac{1}{N^{m/2}\sqrt{s}}\, {\rm tr}\left[{\cal O}(n_{1}) \cdots{\cal O}(n_{m})
\right]\left|0\right\rangle,\:\:\:\:\   m > 1,
\label{state}
\end{equation}
where $\cal O$ represents $A^{\dag}$ or $B^{\dag}$.
The symmetry factor $s$ is the number of possible permutations
of constituents which give the same state \cite{DaKl}.
Note that we should consider only states with two or more constituents
$m > 1$ since we should discard singlet to the leading order of
the $1/N$ expansion of $U(N)$ gauge theory.
It is also absent in the case of $SU(N)$ gauge theory anyway.
All these states satisfy the physical
state condition coming from the constraint (\ref{shiki2})
\begin{equation}
\tilde{J}_{ij}^{+}(0)\left|\Phi\right\rangle = 0.
\end{equation}
Here we note that there are both bosonic and
fermionic oscillators  in our supersymmetric theory.
This fact gives rise to much larger number of new physical states
compared to the purely fermionic or bosonic adjoint matter
case.

Since $P^+$ commutes with other operators, we work on a subspace with
a definite value of the light-cone momentum $P^+$
\begin{equation}
P^{+} = \frac{L}{\pi}K , \quad K= 1, 2, \cdots
\end{equation}
\begin{equation}
K = \sum_{n=1}^{\infty}n\left\{A_{ij}^{\dag}(n)A_{ij}(n)
+B_{ij}^{\dag}(n)B_{ij}(n) \right\} .
\end{equation}
For the state defined in (\ref{state}), $K = \sum_{i=1}^{m}n_{i}$.
Therefore the number of physical states is finite for a given $K$.
So long as $K$ is finite, we can consider finite dimensional
physical state space to diagonalize the mass matrix.
The parameter $K$ plays the role of the infrared cut-off.
The infinite volume limit $L \rightarrow \infty$ is achieved by taking
the limit $K \rightarrow \infty$ with
finite physical values of $P^+$  fixed.
As usual in the discretized light-cone approach, we shall evaluate
mass spectra for finite $K$ corresponding to a finite spatial box
and try to evaluate the asymptotic behavior $K\rightarrow \infty$.

The supersymmetry algebra (\ref{superalgebratwo}) implies that
the diagonalization of the supercharge $Q_2$ gives the
desired mass spectrum.
Let us consider the subspace for fixed light-cone momenta $P^+$, and
denote
\begin{equation}
Q_1=\left[
\begin{array}{cc}
0 &  A^\dagger \\
A &  0
\end{array}
\right], \quad
Q_2=\left[
\begin{array}{cc}
0 &  B^\dagger \\
B &  0
\end{array}
\right] ,
\label{secondsuperchargematrix}
\end{equation}
where the first half of the rows and columns correspond to the bosonic
color singlet bound states and the second half to the fermionic states.
The mass matrix is
\begin{equation}
M^2\equiv 2P^+P^-={\sqrt2 \pi K \over L}
\left[
\begin{array}{cc}
B^\dagger B &  0 \\
0           & B B^\dagger
\end{array}
\right] .
\end{equation}
The diagonalization of the positive definite matrix $B^\dagger B$
gives the mass eigenstates of bosonic color singlet bound states
and the other positive definite matrix $B B^\dagger$ gives
fermionic ones. There exist two unitary matrices $U$ and $V$ such that
\begin{equation}
U^{-1}(B^\dagger B)U=
V^{-1}(B B^\dagger)V=D,
 \quad U^\dagger U=V^\dagger V=1 .
\end{equation}
where the matrix $D$ is positive diagonal.
Let us emphasize that the positive definiteness of mass squared matrices
$B^\dagger B$ and $B B^\dagger$ is a direct consequence of
regularizing the supercharge $Q$ instead of $P^-$ .

The relation (\ref{superalgebraone}) shows that the matrix $A$ is
unitary apart from a scale factor
\begin{equation}
\tilde A \tilde A^\dagger =1, \qquad
\tilde A \equiv 2^{1 \over 4}\sqrt{\pi K \over L}A ,
\end{equation}
The anticommutation relation between two supercharges
(\ref{superalgebrathree}) gives
\begin{equation}
B^\dagger=-\tilde A^\dagger B \tilde A^\dagger ,
\end{equation}
Therefore we find that the matrix $\tilde A$ is precisely the matrix
which maps the mass eigenstates of bosonic bound states and fermionic
ones.
\begin{equation}
V=\tilde A U .
\label{superrelunitarymat}
\end{equation}

In the rest of this section, we consider adding
(supersymmetry-breaking) mass terms $m_b$ for the adjoint scalar field
and $m_f$ for spinor field to explore supersymmetry breaking and
to help treat the zero modes more precisely
\begin{equation}
S_{massive} =S+ \int d^2 x \, {\rm tr} \left[-\frac{1}{2}m_{b}^{2}\phi^{2}
-m_{f}\bar \Psi \Psi \right] ,
\label{massactn}
\end{equation}
where $S$ is the massless action given in eq.(\ref{eq:sacda}).
In the light-cone gauge $A_{-}=0$, the action reduces to mass terms
added to the massless action $S$ in eq.(\ref{eq:action})
\begin{equation}
S_{massive}
= S + \int dx^{+}dx^{-} \, {\rm tr}
\left[
-\frac{1}{2}m_{b}^{2}\phi^{2}
-i\sqrt{2}m_{f}\chi\psi\right].
\label{eq:actionlcgauge}
\end{equation}
The Euler-Lagrange equation for the auxiliary field $\chi$ is modified
from eq.(\ref{eq:const})
\begin{equation}
i\sqrt{2} \partial_{-}\chi
- g[\phi,\psi] -im_f \psi= 0 .
\end{equation}
Elimination of $A_+$ and $\chi$ gives the action with $S$
in eq.(\ref{eq:actionphys}) and mass terms
\begin{equation}
S_{massive} =S+ \int dx^{+}dx^{-} \, {\rm tr}
\left[-\frac{1}{2}m_{b}^{2} \phi^{2}+\frac{i}{2}m_{f}^{2}\psi
\frac{1}{\partial_{-}}\psi
+m_{f}g\psi \frac{1}{\partial_{-}} [\phi , \psi]\right].
\label{eq:actiondayo}
\end{equation}

The momentum $P_{massive}^+$ is the same as eq.(\ref{eq:p+}) and the
Hamiltonian $P_{massive}^{-}$ has mass terms in addition to the
massless $P^-$ in eq.(\ref{eq:p-})
\begin{equation}
P_{massive}^{-} =P^- +  P_{m,quad}^- + P_{m,cubic}^{-}
\end{equation}
\begin{equation}
P^{-}_{m,quad} = \int dx^{-} Tr \Bigg[ \frac{1}{2} m_{b}^{2} \phi^{2}
-\frac{i}{2}m_{f}^{2} \psi \frac{1}{\partial_{-}}\psi \Bigg],
\end{equation}
\begin{equation}
P^{-}_{m,cubic}
=
- m_{f}g\int dx^{-} Tr \left( \psi \frac{1}{\partial_{-}}
[\phi , \psi] \right) .
\label{eq:pv}
\end{equation}

The final result for the additional terms in the Hamiltonian is given
in the discretized light-cone quantization
\begin{equation}
P^{-}_{m,quad}  = \frac{L}{2\pi}  \sum _{n=1}^{\infty}\frac{1}{n}
\left\{m_{b}^{2}A_{ij}^{\dag}(n)A_{ij}(n)
+ m_{f}^{2} B_{ij}^{\dag}(n)B_{ij}(n) \right\},
\end{equation}
\begin{eqnarray}
&\!\!\!
&\!\!\!
 P^{-}_{m,cubic}
 =
-i \frac{m_{f}gL}{4\pi^{3 \over 2}}\sum_{n_1,n_2,n_3=1}^{\infty}
\Bigg\{
\frac{1}{\sqrt{n_2}}\left[\frac{1}{n_3}+\frac{1}{n_1}\right]
\delta_{n_1 -n_2 +n_3,0} A_{kj}^{\dag}(n_2)B_{ki}(n_1)B_{ij}(n_3)
\nonumber\\
&\!\!\!
+
&\!\!\!
\frac{1}{\sqrt{n_1}}
\left[\frac{1}{n_3}-\frac{1}{n_2}\right]\delta_{n_1 -n_2 +n_3,0}
 \left[ A_{jk}(n_1)B^{\dag}_{ik}(n_2)B_{ij}(n_3)-
       A_{ki}(n_1)B^{\dag}_{kj}(n_2)B_{ij}(n_3) \right]\nonumber\\
&\!\!\!
+
&\!\!\!
 \frac{1}{\sqrt{n_1}}\left[\frac{1}{n_3}-\frac{1}{n_2}\right]
\delta_{n_1 +n_2 -n_3,0}
     \left[ A_{kj}^{\dag}(n_1)B_{ik}^{\dag}(n_2)B_{ij}(n_3)-
     A_{ik}^{\dag}(n_1)B_{kj}^{\dag}(n_2)B_{ij}(n_3) \right]\nonumber\\
&\!\!\!
+
&\!\!\!
 \frac{1}{\sqrt{n_1}}\left[\frac{1}{n_2}+\frac{1}{n_3}\right]
\delta_{n_1 -n_2 -n_3,0}
A_{jk}(n_1)B_{ji}^{\dag}(n_2)B_{ik}^{\dag}(n_3)
         \Bigg\}.
\end{eqnarray}

%
%
%%%%%%%  Section 4  %%%%%%%%%%%%%%%%%%%%%%%%%%%%%%%%%%%%%%%%
%
\vspace{7mm}
\pagebreak[3]
\addtocounter{section}{1}
\setcounter{equation}{0}
\setcounter{subsection}{0}
\setcounter{footnote}{0}
\begin{center}
{\large {\bf \thesection. Results of Supercharge Diagonalization}}
\end{center}
\nopagebreak
\medskip
\nopagebreak
\hspace{3mm}

As we have seen, our procedure preserves supersymmetry manifestly
throughout the calculation.
Therefore we are naturally led to obtain supersymmetric mass
spectra with exactly the same bosonic and fermionic spectra for
color singlet states.

If we consider the states with finite values of the discrete momentum
$K$, we have only finitely many physical states to diagonalize
the mass matrix.
Let us illustrate the procedure for smaller values of the
discrete momentum $K$.
In the case of $K=3$,
we find four possible states for bosonic
color singlet states
\begin{equation}
\begin{array}{lcl}
\left|1\right\rangle_{b}&=&
\frac{1}{N^{3/2}\sqrt{3}}\, {\rm tr}\left[A^{\dagger}(1)
A^{\dagger}(1)A^{\dagger}(1)\right]\left|0\right\rangle,\\
\left|2\right\rangle_{b}&=&
\frac{1}{N^{3/2}}\, {\rm tr}\left[A^{\dagger}(1)
B^{\dagger}(1)B^{\dagger}(1)\right]\left|0\right\rangle,\\
\left|3\right\rangle_{b}&=&
\frac{1}{N}\, {\rm tr}\left[A^{\dagger}(2)A^{\dagger}(1)
\right]\left|0\right\rangle,\\
\left|4\right\rangle_{b}&=&\frac{1}{N^{1/2}}\, {\rm tr}\left[B^{\dagger}(2)
B^{\dagger}(1)\right]\left|0\right\rangle.
\label{k4boson}
\end{array}
\end{equation}
and four possible states for fermionic color singlet states
\begin{equation}
\begin{array}{lcl}
\left|1\right\rangle_{f}&=&
\frac{1}{N^{3/2}}\, {\rm tr}\left[A^{\dagger}(1)
A^{\dagger}(1)B^{\dagger}(1)\right]\left|0\right\rangle,\\
\left|2\right\rangle_{f}&=&
\frac{1}{N^{3/2}\sqrt{3}}\, {\rm tr}\left[B^{\dagger}(1)
B^{\dagger}(1)B^{\dagger}(1)\right]\left|0\right\rangle,\\
\left|3\right\rangle_{f}&=&\frac{1}{N}\, {\rm tr}\left[A^{\dagger}(2)
B^{\dagger}(1)
\right]\left|0\right\rangle,\\
\left|4\right\rangle_{f}&=&\frac{1}{N^{1/2}}\, {\rm tr}\left[B^{\dagger}(2)
A^{\dagger}(1)\right]\left|0\right\rangle.
\label{k4fermion}
\end{array}
\end{equation}
Using the matrix $B$ appearing in eq.(\ref{secondsuperchargematrix}),
the mass matrix is given as
\begin{equation}
{\pi \over g^2N}M^2={\pi 2P^+P^- \over g^2N}
={\sqrt2\pi^2 K \over g^2NL}Q_2^2, \quad
Q_2=\left[
\begin{array}{cc}
0 &  B^\dagger \\
B &  0
\end{array}
\right] ,
\end{equation}
\begin{equation}
i{2^{1/4} \pi \sqrt3 \over g\sqrt{NL}}\left(B_{ij}\right)
 = \left(
\begin{array}{cccc}
0&                0  &0&           0 \\
0&                0  &0&-{9 \over 2} \\
0&-3\sqrt{3 \over 2} &0&           0 \\
0&{3 \over 2}\sqrt{3}&0&           0 \\
\end{array}
\right) .
\end{equation}
By diagonalizing this matrix, we obtain mass eigenvalues in units
of $g\sqrt{N/\pi}$.
We find that two bosonic massless states correspond to two fermionic
states through the first supercharge $Q_1$ as shown in
eq.(\ref{superrelunitarymat})
\begin{eqnarray}
\left|1\right\rangle_{b} & \leftrightarrow & \left|1\right\rangle_{f}
\nonumber \\[.5mm]
\left|3\right\rangle_{b} & \leftrightarrow &
{1 \over \sqrt3}\left|3\right\rangle_{f}
+\sqrt{2 \over 3}\left|4\right\rangle_{f}.
\end{eqnarray}
We also find that the two bosonic states with mass eigenvalues
$81/4$ in units of $g^2N/\pi$ correspond to two fermionic states
with the same eigenvalues which are also mapped by the
 first supercharge $Q_1$
\begin{eqnarray}
\left|2\right\rangle_{b} & \leftrightarrow & \left|2\right\rangle_{f}
\nonumber \\[.5mm]
\left|4\right\rangle_{b} & \leftrightarrow &
\sqrt{2 \over 3}\left|3\right\rangle_{f}
-{1 \over \sqrt3}\left|4\right\rangle_{f}.
\end{eqnarray}

Let us note that the adjoint scalar field alone gives only
$\left|1\right\rangle_{b}$ and $\left|3\right\rangle_{b}$ as
color singlet states, whereas the adjoint spinor field alone gives
$\left|4\right\rangle_{b}$ as bosonic color singlet state and
$\left|2\right\rangle_{f}$ as fermionic color singlet state.
Therefore each case gives only a quarter of the possible states
in our supersymmetric theory.

Similarly for $K=4$,
we find nine possible states for bosonic
color singlet states
\begin{equation}
\begin{array}{lcl}
\left|1\right\rangle_{b}&=&{1 \over N^22}
\, {\rm tr}\left[A^{\dagger}(1)A^{\dagger}(1)A^{\dagger}(1)
A^{\dagger}(1)\right]\left|0\right\rangle\\
\left|2\right\rangle_{b}&=&{1 \over N^2}
\, {\rm tr}\left[A^{\dagger}(1)A^{\dagger}(1)
B^{\dagger}(1)B^{\dagger}(1)\right]\left|0\right\rangle\\
\left|3\right\rangle_{b}&=&{1 \over N^{3/2}}
\, {\rm tr}\left[A^{\dagger}(2)A^{\dagger}(1)
A^{\dagger}(1)\right]\left|0\right\rangle\\
\left|4\right\rangle_{b}&=&{1 \over N^{3/2}}
\, {\rm tr}\left[A^{\dagger}(2)B^{\dagger}(1)
B^{\dagger}(1)\right]\left|0\right\rangle\\
\left|5\right\rangle_{b}&=&{1 \over N^{3/2}}
\, {\rm tr}\left[A^{\dagger}(1)B^{\dagger}(1)
B^{\dagger}(2)\right]\left|0\right\rangle\\
\left|6\right\rangle_{b}&=&{1 \over N^{3/2}}
\, {\rm tr}\left[A^{\dagger}(1)B^{\dagger}(2)
B^{\dagger}(1)\right]\left|0\right\rangle\\
\left|7\right\rangle_{b}&=&{1 \over N}
\, {\rm tr}\left[A^{\dagger}(3)A^{\dagger}(1)\right]\left|0\right\rangle\\
\left|8\right\rangle_{b}&=&{1 \over N\sqrt{2}}\, {\rm tr}\left[A^{\dagger}(2)
A^{\dagger}(2)\right]\left|0\right\rangle\\
\left|9\right\rangle_{b}&=&{1 \over N}
\, {\rm tr}\left[B^{\dagger}(3)B^{\dagger}(1)\right]\left|0\right\rangle ,
\end{array}
\end{equation}
and nine possible states for fermionic color singlet states
\begin{equation}
\begin{array}{lcl}
\left|1\right\rangle_{f}&=&{1 \over N^2}
\, {\rm tr}\left[A^{\dagger}(1)A^{\dagger}(1)
A^{\dagger}(1)B^{\dagger}(1)\right]\left|0\right\rangle\\
\left|2\right\rangle_{f}&=&{1 \over N^2}
\, {\rm tr}\left[A^{\dagger}(1)B^{\dagger}(1)
B^{\dagger}(1)B^{\dagger}(1)\right]\left|0\right\rangle\\
\left|3\right\rangle_{f}&=&{1 \over N^{3/2}}
\, {\rm tr}\left[A^{\dagger}(2)A^{\dagger}(1)
B^{\dagger}(1)\right]\left|0\right\rangle\\
\left|4\right\rangle_{f}&=&{1 \over N^{3/2}}
\, {\rm tr}\left[A^{\dagger}(2)B^{\dagger}(1)
A^{\dagger}(1)\right]\left|0\right\rangle\\
\left|5\right\rangle_{f}&=&{1 \over N^{3/2}}
\, {\rm tr}\left[B^{\dagger}(2)A^{\dagger}(1)
A^{\dagger}(1)\right]\left|0\right\rangle\\
\left|6\right\rangle_{f}&=&{1 \over N^{3/2}}
\, {\rm tr}\left[B^{\dagger}(2)B^{\dagger}(1)
B^{\dagger}(1)\right]\left|0\right\rangle\\
\left|7\right\rangle_{f}&=&{1 \over N}
\, {\rm tr}\left[A^{\dagger}(3)B^{\dagger}(1)\right]\left|0\right\rangle\\
\left|8\right\rangle_{f}&=&{1 \over N}
\, {\rm tr}\left[A^{\dagger}(2)B^{\dagger}(2)\right]\left|0\right\rangle\\
\left|9\right\rangle_{f}&=&{1 \over N}
\, {\rm tr}\left[B^{\dagger}(3)A^{\dagger}(1)\right]\left|0\right\rangle .
\end{array}
\end{equation}
The matrix $B$ appearing in the supercharge $Q_2$
 in eq.(\ref{secondsuperchargematrix})
is given by
\begin{equation}
i\frac{ 2^{5/4}\pi}{g\sqrt{NL}}\left(B_{ij}\right) = \left(
\begin{array}{ccccccccc}
0&0&0&0&0&0&0&0&0 \\
0&0&0&0&3&-3&0&0&0 \\
0&-{3 \over \sqrt{2}}&0&0&0&0&-{5 \over \sqrt{6}}&3
&{1 \over 3\sqrt{2}} \\
0&-{3 \over \sqrt{2}}&0&0&0&0&{5 \over \sqrt{6}}&-3
&-{1 \over 3\sqrt{2}} \\
0&3&0&0&0&0&0&0&0 \\
0&0&0&0&0&0&0&0&-{14 \over 3} \\
0&0&0&-5\sqrt{\frac{2}{3}}&-{2 \over \sqrt{3}}
&-{2 \over \sqrt{3}}&0&0&0 \\
0&0&0&3&-{3 \over \sqrt{2}}&-{3 \over \sqrt{2}}&0&0&0 \\
0&0&0&0&{7 \over 3}&{7 \over 3}&0&0&0 \\
\end{array}
\right) .
\label{K4B}
\end{equation}

 From the diagonalization of the matrix
 for bosonic color singlet states,
we find four different mass eigenvalues
$0$,  $18$, and $(1302 \pm 42\sqrt{13})/ 54$ .
All massive states have degeneracy two, whereas there are three
massless states.
We find exactly the same spectra for fermionic color singlet states.

We have explicitly constructed bosonic and fermionic color singlet
states for higher values of the cut-off momentum $K$ up to $K=11$.
We find the number of bosonic color singlet states for
$K=5, 6, 7, 8, 9, 10,$ and $ 11$ to be $24$, $61$, $156$, $409$, $1096$,
$2953$, and $ 8052$ respectively.
The number of fermionic color singlet states is exactly the same
as the corresponding bosonic one with the same $K$.

After evaluating the supercharge for these subspace up to $K=8$,
we diagonalize the supercharge exactly to obtain the mass
eigenvalues.
In Fig.\ref{rui} we plot the accumulated number of bosonic color
singlet bound states as a function of mass squared in units of
$g^2 N/\pi$.
We can see that the number of states is approaching to a limiting value
at least for smaller values of $M^2$.
The present tendency seems to suggest that the density of states is
increasing rapidly as the mass squared increases.
This behavior is in qualitative agreement with
the previous results
for the adjoint scalar or adjoint spinor matter constituents in
nonsupersymmetric gauge theories \cite{DaKl}.
Namely the density of states showed an exponential increase
as mass squared increases in accordance with the closed string
interpretation.
The fermionic color singlet bound states show the same behavior.

In Fig.\ref{k5fig}, %\ref{k8ffig},
we plot the mass squared of
bosonic
color singlet bound states
in units of $g^2N/\pi$
as a function of the average number of constituents
for the case of $K=5$.
We have also obtained a similar plot of the fermionic color singlet
bound states which turns out to be indistinguishable from the bosonic
one.
Since we find the exact correspondence, we shall display only
the bosonic spectra.
In Figs.\ref{k6fig}, \ref{k7fig}, and \ref{k8bfig}
we plot the mass squared in units of $g^2N/\pi$
as a function of the average number of constituents for the
case of $K=6, 7,$ and $ 8$ respectively.
It is interesting to see that the average number of constituents
increases as mass squared increases.

We find that there are a number of massless states.
Empirically we find that there are $K-1$ bosonic and fermionic
massless states for the momentum cut-off $K$.
It is easy to understand some of the massless states.
For instance, for each $K$ there is one massless bosonic state with
$K$ bosonic oscillators of the lowest level $A^\dagger(1)$ acting
on the vacuum.
There is also one massless bosonic state with one bosonic
oscillator $A^\dagger(2)$ of level two and $K-2$ bosonic
oscillators of the lowest level $A^\dagger(1)$ acting on the vacuum.
Both these states become massless at arbitrary $K$
because of the principal value
prescription for the infrared regularization of the supercharge.

The bound state equations for adjoint scalar or spinor constituents
are infinitely coupled even in the large $N$ limit \cite{DaKl}.
To compare with the case of constituents in the fundamental
representation, it is instructive to work out a truncation to a two
constituents subspace.
The two-body bound state equation becomes analogous to but is somewhat
 different from the 'tHooft equation \cite{THooft} extended to the
boson-boson bound state case \cite{Shei} and the boson-fermion
bound state case \cite{Aoki}, as given in Appendix B.
Unfortunately, our results in Figs.\ref{k5fig}--\ref{k8bfig}
suggest that the two-body truncation does not seem to give an adequate
approximation even for states with low excitations.

To explore the effects of supersymmetry breaking mass terms,
we diagonalize the mass matrix exactly
with equal mass $m=m_b=m_f$ for scalar and spinor
constituents.
The explicit form of the mass matrix for $K=3$ and $K=4$
are given in Appendix C.
As an illustration, we plot the mass squared of bosonic color singlet
bound states for $K=4$ as a function of the constituent mass
both in unit of $g\sqrt{N}/\pi$ in Fig.\ref{k4bmm}.
Similarly Fig.\ref{k4fmm} shows the fermionic bound state.
We observe that the mass spectra of bosonic
bound states and fermionic ones
differ as constituent mass increases even though
we have given identical masses for both bosonic and fermionic
constituents.
This is because they are supersymmetric partners of gauge boson
which has to be massless.
It is interesting to see that the vanishing mass of the
gauge boson demands massless scalars and spinors even though the
gauge boson does not have dynamical degree of freedom.

%
%%%%%%% %%%%%%%%%%%%%%%%%%%%%%%%%%%%%%%%%%%%%%%%
%
\vspace{5mm}
%
%%%%%%%  Acknowledgement  %%%%%%%%%%%%%%%%%%%%%%%%%%%%%%%%%%%
%

We wish to acknowledge Simon Dalley for a useful discussion and
advise on diagonalization of matrices, Kenichiro Aoki for
an illuminating discussion and Des Johnston for a reading of
the manuscript.
One of the authors (N.S) thanks Tohru Eguchi, Kiyoshi Higashijima,
Sung-Kil Yang, and Elcio and Christina Abdalla for an interesting
discussion.
One of the authors (N.S.) would like to thank the Aspen Center for
Physics and Service de Physique Theorique Saclay for hospitality,
and the Japan Society for the Promotion of Science for a grant.
This work is supported in part by Grant-in-Aid for
Scientific Research (No.05640334), and
Grant-in-Aid for Scientific Research for Priority Areas
(No.05230019) from the Ministry of Education, Science
and Culture.

\vspace{5mm}
%
%%%%%%%  Appendix A  %%%%%%%%%%%%%%%%%%%%%%%%%%%%%%%%%%%%%%%%%%
%
%\def\theequation{A.\arabic{equation}}
\vspace{7mm}
\pagebreak[3]
\setcounter{section}{1}
\setcounter{equation}{0}
\setcounter{subsection}{0}
\setcounter{footnote}{0}
\begin{center}
{\large{\bf Appendix A.
Superfield and Supertransformation
}}
\end{center}
\nopagebreak
\medskip
\nopagebreak
\hspace{3mm}

Here we construct the action of supersymmetric Yang-Mills theory in
$1+1$ dimensions by using the superfield formalism.

The spinor superfield $V_\alpha$ corresponds
to the vector multiplet
\begin{equation}
V_{\alpha}(x , \theta) = \xi_{\alpha}(x) -
\frac{i}{2}(\gamma_{5}\gamma_{\mu}\theta)_{\alpha} \frac{A^{\mu}(x)}{g}
+ \frac{1}{2} \theta_{\alpha} \phi(x) - \frac{1}{2}
N(x)(\gamma_{5} \theta)_{\alpha} -
\frac{1}{2} \bar{\theta}\theta \sqrt{2}\Psi_{\alpha}(x) ,
\label{eq:sv}
\end{equation}
where $\theta$ is a two-component Majorana Grassmann spinor,
$\xi_{\alpha} , \Psi_{\alpha}$ are Majorana spinors, $A^{\mu}$ is
a vector field, $\psi$ and $N$ are scalar fields.
Spinor indices and spacetime indices are
denoted by $\alpha$ and $\mu$ respectively.
The infinitesimal gauge transformation on $V_{\alpha}$
is defined by
\begin{equation}
\delta_{gauge}V_{\alpha} =
-(\gamma_{5}D)_{\alpha}S - i2g [S , V_{\alpha}]
=-(\gamma_5\nabla)_\alpha S ,
\label{eq:nabegt}
\end{equation}
\begin{equation}
(\nabla)_\alpha S
\equiv
D_{\alpha}S - i2g [(\gamma_5V)_{\alpha}, S] ,
\quad
D_{\alpha} = - \frac{\partial}{\partial \bar{\theta^{\alpha}}} +
i(\gamma^{\mu}\theta)_{\alpha} \frac{\partial}{\partial x^{\mu}} ,
\end{equation}
where
$D_{\alpha}$ is the supercovariant derivative and
$\nabla$ is the super- as well as gauge- covariant derivative.
The transformation parameter $S$ is
a scalar superfield:
\begin{equation}
S(x , \theta) = \Lambda(x) - \bar{\theta}\lambda(x) -
\frac{1}{2}\bar{\theta}\theta F(x) ,
\label{eq:ss}
\end{equation}
where $\Lambda , F$ are scalar fields, and $\lambda$
is a two-component Majorana spinor field.

Let us define the quantity $\tilde G$ which transforms covariantly
under the gauge transformation (\ref{eq:nabegt})
\begin{equation}
\tilde{G} =\bar{D}V + i2g \bar{V} \gamma_{5} V ,
\quad
\delta_{gauge} \tilde{G} = -i2g[S , \tilde{G}] .
\label{eq:tilggt}
\end{equation}
Using $\int d^2 \theta {1 \over 2}\bar{\theta}\theta = -1$,
the action of supersymmetric Yang-Mills theory is given by the
covariant derivative of $\tilde G$
\begin{equation}
\nabla_{\alpha} \tilde{G} = D_{\alpha}\tilde{G}- i2g[\gamma_{5}V ,
\tilde{G}] ,
\quad
\delta_{gauge}\nabla_{\alpha} \tilde{G}
= - i2g[S , \nabla_{\alpha}\tilde{G}] .
\end{equation}
\begin{equation}
 S =
\int d^{2}\theta d^{2}x \Bigg[-\frac{1}{4} \, {\rm tr}
\left(\bar{\nabla}_{\alpha}\tilde{G}\nabla_{\alpha}
\tilde{G}\right)\Bigg] .
\label{eq:sac}
\end{equation}
Let us decompose the gauge transformation (\ref{eq:nabegt})
 in component fields
\begin{equation}
\begin{array}{lcl}
\delta_{gauge}\xi_{\alpha} &=& -(\gamma_{5} \lambda)_{\alpha}
-i2g[\Lambda , \xi_{\alpha} ] , \\
\displaystyle\delta_{gauge} A_{\mu}/g
&=& -i2[\Lambda ,A_{\mu}] - 2g
\{ \bar{\lambda} , \gamma_{5} \gamma_{\mu} \xi \}
+ 2\partial_{\mu}\Lambda , \\
\delta_{gauge} \phi &=& -i2g[\Lambda , \phi]
-i2g\{\bar{\lambda} , \xi\} , \\
\delta_{gauge}N &=& 2F - i2g[\Lambda , N] + i2g \{\bar{\lambda} ,
\gamma_{5}\xi\}, \\
\delta_{gauge}\sqrt{2}\Psi &=&
i\gamma_{5}\gamma_{\mu}\partial^{\mu}\lambda
-i2g[\Lambda , \sqrt{2}\Psi]
+ [\gamma_{5}\gamma^{\mu}\lambda , A_{\mu}] \\
  & & +ig[\lambda , \phi] -ig[\gamma_{5}\lambda , N] - i2g[F , \xi] .
\end{array}
\label{comgaugetr}
\end{equation}
We choose the Wess-Zumino gauge $\xi_{\alpha} = N = 0$ by using
the gauge freedom $\lambda$ and $F$, and we find the remaining gauge
transformation with the parameter $\Lambda$
\begin{eqnarray}
 \delta_{gauge} A_{\mu}
&\!\!\!
=&\!\!\!
 -i2g[\Lambda ,A_{\mu}]
+ 2g \partial_{\mu}\Lambda, \\
\delta_{gauge} \phi
&\!\!\!=&\!\!\! -i2g[\Lambda , \phi], \quad
\delta_{gauge}\sqrt{2}\Psi = -i2g[\Lambda , \sqrt{2}\Psi] . \nonumber
\end{eqnarray}

Next we consider the supertransformation. The superfield $V$ transforms
as
\begin{equation}
\delta_{super}V
= -i\bar{\epsilon}Q V = \bar{\epsilon}\left[-\frac{\partial}
{\partial\bar{\theta}} - i\gamma^{\mu}\theta \partial_{\mu} \right] V ,
\end{equation}
where $Q_{\alpha}$ is the supercharge acting on superfields
and $\epsilon$ is an infinitesimal two-component Majorana spinor.
In terms of components, it becomes
\begin{equation}
\begin{array}{lcl}
\delta_{super}\xi_{\alpha} &=& \displaystyle \frac{i}{2}
(\gamma_{5}\gamma^{\mu}\epsilon)_{\alpha}\frac{A_{\mu}}{g}
-\frac{1}{2}\epsilon_{\alpha}\phi +{1 \over 2}\left(\gamma_{5}\epsilon
\right)_{\alpha}N , \\
\delta_{super}A_{\mu} &=&
ig\bar{\epsilon}\gamma_{5}\left(\gamma_{\mu}\sqrt{2}\Psi
+i\gamma^{\nu}\gamma_{\mu}\partial_{\nu}\xi\right) , \\
\delta_{super}\phi &=& -\bar{\epsilon}\left(\sqrt{2}\Psi
-i\gamma^{\nu}\partial_{\nu}\xi\right) , \\
\delta_{super}N &=& \bar{\epsilon} \gamma_{5}\left(\sqrt{2}\Psi
+i\gamma^{\nu}\partial_{\nu}\xi\right) , \\
\delta_{super}\Psi_{\alpha} &=& \displaystyle \frac{1}{2}(\gamma_{5}
\gamma^{\nu}\gamma^{\mu}\epsilon)_{\alpha}
\frac{\partial_{\mu}A_{\nu}}{g}
+ \frac{i}{2}(\gamma^{\mu}\epsilon)_{\alpha}\partial_{\mu}\phi
-{i \over 2}\left(\gamma_{5}\gamma^{\mu}\epsilon\right)_{\alpha}
\partial_{\mu}N .
\end{array}
\end{equation}
Note that the Wess-Zumino gauge condition $\xi = N = 0$ is violated
by the supertransformation.
We  therefore need to make compensating gauge transformation to maintain
the Wess-Zumino gauge condition
\begin{equation}
\delta_{super}\xi + \delta_{gauge}\xi = 0 ,\:\:
\delta_{super}N + \delta_{gauge}N = 0,
\end{equation}
at the Wess-Zumino gauge fixing slice $\xi = N = 0$.
By choosing the compensating gauge transformation as
\begin{equation}
\lambda = \displaystyle \frac{i}{2}\gamma^{\mu}\epsilon\frac{A_{\mu}}{g}
- \frac{1}{2}\gamma_{5}\epsilon\phi, \quad
F = \displaystyle -\frac{1}{2}\bar{\epsilon}\gamma_{5}\sqrt{2}\Psi .
\end{equation}
we obtain the modified supertransformation
$\tilde{\delta}_{super}\equiv \delta_{super}+\delta_{gauge}$
in the Wess-Zumino gauge as shown in eq.(\ref{eq:st}).
In the Wess-Zumino gauge, $\tilde{G}$ becomes
\begin{equation}
\tilde{G} = \phi + \bar{\theta}\sqrt{2} \Psi + \frac{1}{2}
\bar{\theta}\theta\frac{1}{2g}\epsilon^{\mu\nu}F_{\mu\nu} ,
\label{eq:tilg}
\end{equation}
where $\epsilon^{01}=-\epsilon_{01}=1$.
Substituting (\ref{eq:tilg}) into (\ref{eq:sac}),
we obtain the action (\ref{eq:sacda})
of the two dimensional supersymmetric Yang-Mills theory.
Since the light-cone gauge condtion $A_-=0$ is violated by the
supertransformation (\ref{eq:st}), we need to define a modified
supertransformation
$\tilde{\tilde \delta}_{super}\equiv\tilde \delta_{super}+
\delta_{gauge}$ by adding another
compensating gauge transformation to make
$\tilde{\tilde \delta}_{super} A_-=0$.
We find the supertransformation for the dynamical variables
in the light-cone gauge as
\begin{equation}
\tilde{\tilde \delta}_{super} \phi
=
i 2^{1/4} \left(\epsilon_1 \chi-\epsilon_2 \psi\right) +
 2^{3/4} g \epsilon_1 \left[{1 \over \partial_-}\psi, \phi\right] \\
\label{lightconesuperphi}
\end{equation}

\begin{equation}
\tilde{\tilde \delta}_{super} \psi
=
- {1 \over 2^{1/4}g} \epsilon_1 \partial_- A_+
+ 2^{1/4}\epsilon_2 \partial_-\phi +
 2^{3/4} g \epsilon_1 \left[{1 \over \partial_-}\psi, \psi\right]
\label{lightconesuperpsi}
\end{equation}
%
%%%%%%%  Appendix B  %%%%%%%%%%%%%%%%%%%%%%%%%%%%%%%%%%%%%%%%%%
%
%\def\theequation{B.\arabic{equation}}
\vspace{7mm}
\pagebreak[3]
\setcounter{section}{1}
\setcounter{equation}{0}
\setcounter{subsection}{0}
\setcounter{footnote}{0}
\begin{center}
{\large{\bf Appendix B.
Tow-body Truncation of Bound State Equations
}}
\end{center}
\nopagebreak
\medskip
\nopagebreak
\hspace{3mm}

Here we summarize the bound state equations in the truncated subspace of
two constituents only.
Bosonic bound states consist of two bosonic consitituents wave
function $\phi_{bb}$ and two fermionic one $\phi_{ff}$
\begin{eqnarray}
\left\vert  \Phi(P^+) \right\rangle_b
&\!\!\!
=
&\!\!\!
\int_0^{P^+} dk_1 dk_2 \delta(k_1+k_2-P^+)
\biggl\{\phi_{bb}(k_1,k_2){1 \over N}
{\rm tr}[a^\dagger(k_1), a^\dagger(k_2)] \nonumber \\
&\!\!\!
+
&\!\!\!
\phi_{ff}(k_1,k_2){1 \over N}
{\rm tr}[b^\dagger(k_1), b^\dagger(k_2)]\biggr\}
\left\vert 0 \right\rangle ,
\end{eqnarray}
\begin{equation}
\phi_{bb}(k, P^+-k)=\phi_{bb}(P^+-k, k),
\;\;\; \phi_{ff}(k, P^+-k)=-\phi_{ff}(P^+-k, k) .
\end{equation}
We define
%\begin{equation}
$C_j=C+{2\pi m_j \over g^2 N}$, %\;
($j=b, f$),
%\end{equation}
using the quadratic term in the
Hamiltonian $C(k)$ defined in eq.(\ref{quadcoefficient}).
We obtain a coupled bound state equation for bosonic bound states
using $x\equiv k/P^+$, $y\equiv l/P^+$
\begin{eqnarray}
M^2\phi_{bb}(k, P^+-k)
&\!\!\!
=
&\!\!\!
{g^2N \over 2\pi}\left[{C_b( k ) \over x} +
{C_b( P^+-k ) \over 1-x}\right]\phi_{bb}(k,P^+-k) \nonumber \\
&\!\!\!
-
&\!\!\!
{g^2N \over 2\pi}\int_0^1
{dy \over \sqrt{x(1-x)y(1-y)}}{(x+y)(2-x-y) \over
(x-y)^2}\phi_{bb}(l, P^+-l) \nonumber \\
&\!\!\!
+
&\!\!\!
 {g^2N \over 2\pi} \int_0^1 {dy \over(y-x)\sqrt{x(1-x)}}
\phi_{ff}(l, P^+-l),
\end{eqnarray}

\begin{eqnarray}
&\!\!\!
&\!\!\!
M^2 \phi_{ff}(k, P^+-k)
= {g^2N \over 2\pi}\left[{C_f(k) \over x}+
{C_f(P^+-k) \over 1-x}\right]\phi_{ff}(k, P^+-k) \nonumber  \\
&\!\!\!
-
&\!\!\!
{2g^2N \over \pi}\int_0^1 {dy \over (x-y)^2}
\phi_{ff}(l, P^+-l) +{g^2N \over 2\pi}\int_0^1 dy
{\phi_{bb}(l, P^+-l) \over (x-y)\sqrt{y(1-y)}} .
\end{eqnarray}

Similarly we find a bound state equation for fermionic ones
\begin{equation}
\left\vert  \Phi(P^+) \right\rangle_f
=
\int_0^{P^+} dk_1 dk_2 \delta(k_1+k_2-P^+)
\phi_{bf}(k_1,k_2){1 \over N}
{\rm tr}[a^\dagger(k_1), b^\dagger(k_2)]
\left\vert  0 \right\rangle ,
\end{equation}
\begin{eqnarray}
&\!\!\!
&\!\!\!
M^2\phi_{bf}(k, P^+-k)
=
{g^2N \over 2\pi}\left[{C_b(k) \over x}+
{C_f(P^+-k) \over 1-x}\right]\phi_{bf}(k, P^+-k) \\
&\!\!\!
-
&\!\!\!
{g^2N \over 2\pi}\int_0^1 {dy \over (1-x-y)\sqrt{xy}}
\phi_{bf}(l, P^+-l)
-
{g^2N \over \pi}\int_0^1 dy {x+y \over (x-y)^2\sqrt{xy}}
\phi_{bf}(l, P^+-l). \nonumber
\end{eqnarray}

%
%%%%%%%  Appendix C  %%%%%%%%%%%%%%%%%%%%%%%%%%%%%%%%%%%%%%%%%%
%
%\def\theequation{C.\arabic{equation}}
\vspace{7mm}
\pagebreak[3]
\setcounter{section}{1}
\setcounter{equation}{0}
\setcounter{subsection}{0}
\setcounter{footnote}{0}
\begin{center}
{\large{\bf Appendix C.
Mass Matrix with Massive Constituents
}}
\end{center}
\nopagebreak
\medskip
\nopagebreak
\hspace{3mm}

Here we display the bound state mass matrices for massive constituents.
Introducing mass squared parameters of constituents
in unit of $g^2 N/\pi$
\begin{equation}
x=\frac{\pi m_{b}^{2}}{g^2N}, \qquad y=\frac{\pi m_{f}^{2}}{g^2N}
\end{equation}
we find the mass squared matrix in unit of $g^2N/\pi$
for $K=3$ bosonic bound states defined in eq.(\ref{k4boson})
\begin{equation}
{M^2 \pi \over g^2N}=
\left(
\begin{array}{cccc}
9x&0&0&0 \\
0&{81 \over 4} + 3x + 6y&-3i\sqrt{{y \over 2}}&-{3i \over 2}\sqrt{y} \\
0&3i\sqrt{{y \over 2}}&{9 \over 2}x&0 \\
0&{3i \over 2}\sqrt{y}&0&{81 \over 4} + {9 \over 2}y \\
\end{array}
\right)
\end{equation}

For $K=3$ fermionic bound states defined in eq.(\ref{k4fermion}), we find
\begin{equation}
{M^2 \pi \over g^2N}=
\left(
\begin{array}{cccc}
3x + 6y&0&0&0 \\
0&{81 \over 4} + 9y&-3i\sqrt{{3y \over 2}}&0 \\
0&3i\sqrt{{3y \over 2}}&{27 \over 2} + {3 \over 2}x + 3y&
   -{27 \over 2\sqrt{2}}\\
0&0&-{27 \over 2\sqrt{2}}&{27 \over 4} + 3x + {3 \over 2}y \\
\end{array}
\right)
\end{equation}

For $K=4$ bound states, we decompose the mass matrix as
\begin{equation}
{\pi \over g^2N}M^2={\sqrt2\pi^2 K \over g^2NL}Q_2^2+T_x+T_y
\end{equation}
For bosonic bound states,
the first supersymmetric term $Q_2^2$ reduces to $B^\dagger B$
in terms of the matrix $B$ in eq.(\ref{K4B}),
and the second and third terms are given by
\begin{equation}
T_x=\left(
\begin{array}{ccccccccc}
16x&0&0&0&0&0&0&0&0 \\
0&8x&0&0&0&0&0&0&0 \\
0&0&10x&0&0&0&0&0&0 \\
0&0&0&2x&0&0&0&0&0 \\
0&0&0&0&4x&0&0&0&0 \\
0&0&0&0&0&4x&0&0&0 \\
0&0&0&0&0&0&{16 \over 3}x&0&0 \\
0&0&0&0&0&0&0&4x&0 \\
0&0&0&0&0&0&0&0&0 \\
\end{array}
\right)
\end{equation}

\begin{equation}
T_y=\left(
\begin{array}{ccccccccc}
0&0&0&0&0&0&0&0&0 \\
0&8y&-2i\sqrt{2y}&0&i\sqrt{y}&-i\sqrt{y}&0&0&0 \\
0&2i\sqrt{2y}&&0&0&0&0&0&0 \\
0&0&0&8y&0&0&0&-4i\sqrt{y}&-{4i \over 3}\sqrt{2y} \\
0&-i\sqrt{y}&0&0&6y&0&-i\sqrt{3y}&0&-{i \over 3}\sqrt{y} \\
0&i\sqrt{y}&0&0&0&6y&-i\sqrt{3y}&0&-{i \over 3}\sqrt{y} \\
0&0&0&0&i\sqrt{3y}&i\sqrt{3y}&0&0&0 \\
0&0&0&4i\sqrt{y}&0&0&0&0&0 \\
0&0&0&{4i \over 3}\sqrt{2y}&{i \over 3}\sqrt{y}&{i \over 3}\sqrt{y}&
   0&0&{16 \over 3}y\\
\end{array}
\right)
\end{equation}
For fermionic bound states,
the $Q_2^2$ reduces to $B B^\dagger$
in terms of the matrix $B$ in eq.(\ref{K4B}),
and the second and third terms are given by
\begin{equation}
T_x=\left(
\begin{array}{ccccccccc}
12x&0&0&0&0&0&0&0&0 \\
0&4x&0&0&0&0&0&0&0 \\
0&0&6x&0&0&0&0&0&0 \\
0&0&0&6x&0&0&0&0&0 \\
0&0&0&0&8x&0&0&0&0 \\
0&0&0&0&0&0&0&0&0 \\
0&0&0&0&0&0&{4 \over 3}x&0&0 \\
0&0&0&0&0&0&0&2x&0 \\
0&0&0&0&0&0&0&0&4x \\
\end{array}
\right)
\end{equation}

\begin{equation}
T_y=\left(
\begin{array}{ccccccccc}
4y&0&0&0&0&0&0&0&0 \\
0&12y&-2i\sqrt{2y}&-2i\sqrt{2y}&0&0&0&0&0 \\
0&2i\sqrt{2y}&4y&0&0&0&0&-i\sqrt{y}&{2i \over 3}\sqrt{2y} \\
0&2i\sqrt{2y}&0&4y&0&0&0&i\sqrt{y}&-{2i \over 3}\sqrt{2y} \\
0&0&0&0&2y&0&0&0&0 \\
0&0&0&0&0&10y&-2i\sqrt{3y}&-2i\sqrt{2y}&0 \\
0&0&0&0&0&2i\sqrt{3y}&4y&0&0 \\
0&0&i\sqrt{y}&-i\sqrt{y}&0&2i\sqrt{2y}&0&2y&0 \\
0&0&-{2i \over 3}\sqrt{2y}&{2i \over 3}\sqrt{2y}&0&0&0&0&{4 \over 3}y \\
\end{array}
\right)
\end{equation}

%%%%%%%  References  %%%%%%%%%%%%%%%%%%%%%%%%%%%%%%%%%%%%%%%
\vspace{5mm}
%\newpage
%
%\newcommand{\NP}[1]{{\it Nucl.\ Phys.\ }{\bf #1}}
%\newcommand{\PL}[1]{{\it Phys.\ Lett.\ }{\bf #1}}
%\newcommand{\CMP}[1]{{\it Commun.\ Math.\ Phys.\ }{\bf #1}}
%\newcommand{\MPL}[1]{{\it Mod.\ Phys.\ Lett.\ }{\bf #1}}
%\newcommand{\IJMP}[1]{{\it Int.\ J. Mod.\ Phys.\ }{\bf #1}}
%\newcommand{\PR}[1]{{\it Phys.\ Rev.\ }{\bf #1}}
%\newcommand{\PRL}[1]{{\it Phys.\ Rev.\ Lett.\ }{\bf #1}}
%\newcommand{\PTP}[1]{{\it Prog.\ Theor.\ Phys.\ }{\bf #1}}
%\newcommand{\PTPS}[1]{{\it Prog.\ Theor.\ Phys.\ Suppl.\ }{\bf #1}}
%\newcommand{\AP}[1]{{\it Ann.\ Phys.\ }{\bf #1}}
%

\def\epsfbox{} %NO FIGS!!!
%
%%%%%%%%%%%% figure caption %%%%%%%%%%%%%%%%%
%
\section*{Figure captions}
\begin{itemize}
\item[Fig.\ 1]
The accumulated number of bound states as a function of mass squared
for $K=4, 5, 6, 7, 8$; there is no differnce in behavior between bosonic
and fermionic state.

\item[Fig.\ 2]
Mass squared of bosonic bound states for $K=5$ as a function of
the average number of constituents; $M^2$ are measured in units of
$g^2 N / \pi$.

\item[Fig.\ 3]
Mass squared of $K=6$ bosonic bound states as a function of
the average number of constituents; $M^2$ are measured in units of
$g^2 N / \pi$.

\item[Fig.\ 4]
Mass squared of $K=7$ bosonic bound states as a function of
the average number of constituents; $M^2$ are measured in units of
$g^2 N / \pi$.

\item[Fig.\ 5]
Mass squared of $K=8$ bosonic bound states as a function of
the average number of constituents; $M^2$ are measured in units of
$g^2 N / \pi$.

\item[Fig.\ 6]
Mass squared of $K=4$ bosonic bound states as a function of
the constituent mass squared; both are measured in units of
$g^2 N / \pi$.
%Mass squared of fermionic bound states for $K=8$ as a function of
%the average number of constituents

\item[Fig.\ 7]
Mass squared of $K=4$ fermionic bound states as a function of
the constituent mass squared; both are measured in units of
$g^2 N / \pi$.
%The number of bound states as a function of mass squared for
%$K=5, 6, 7, 8$
\end{itemize}

%
%%%%%%%%%% figure %%%%%%%%%%%%%%%%%%%%%%%%%%
%
%
%
\begin{figure}
 \leavevmode
 \epsfysize=18cm
 \centerline{\epsfbox{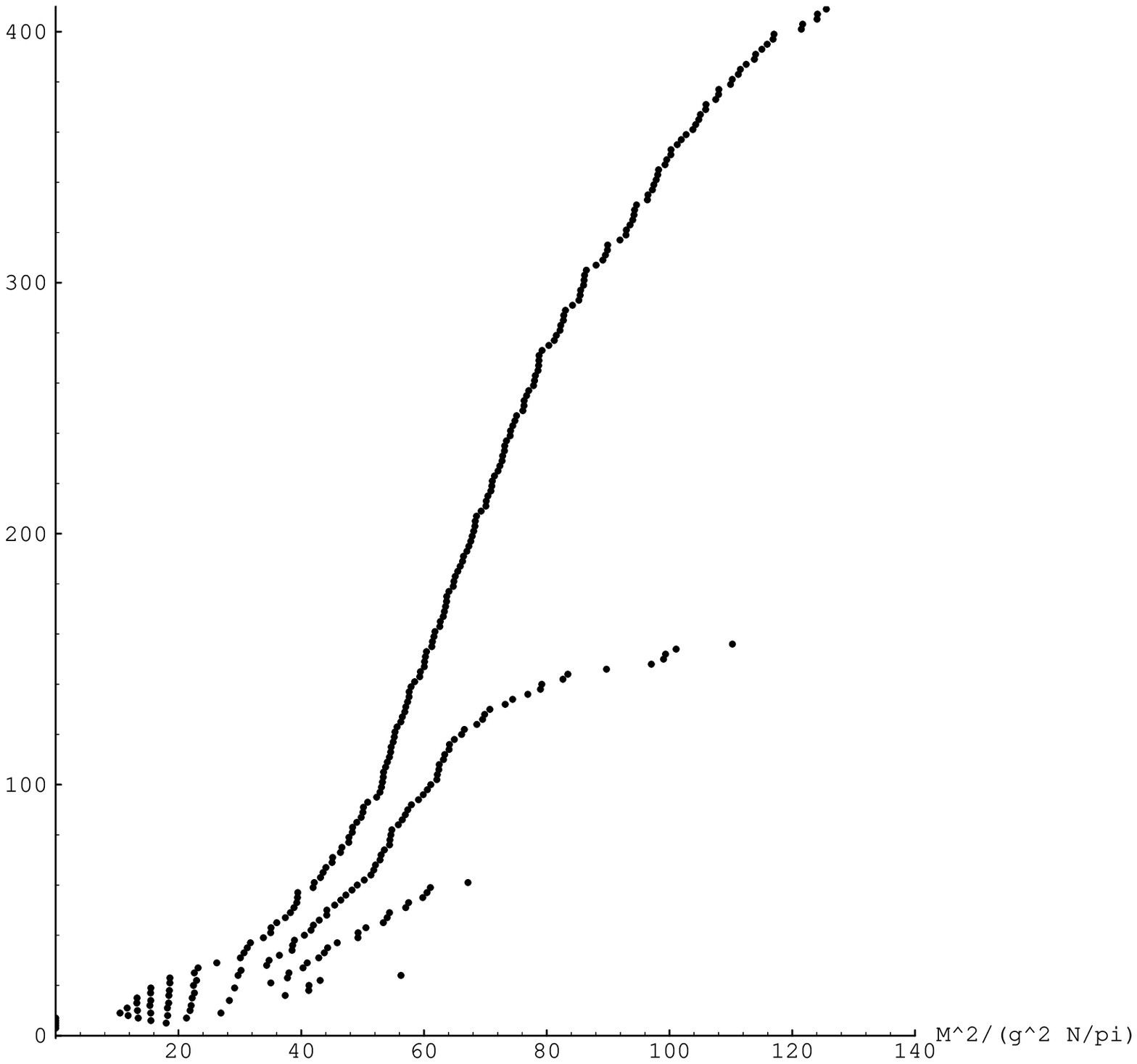}}
 \caption{
The accumulated number of bound states as a function of mass squared
for $K=4, 5, 6, 7, 8$; there is no differnce in behavior between bosonic
and fermionic state.
}
 \label{rui}
\end{figure}

\begin{figure}
 \leavevmode
 \epsfysize=18cm
 \centerline{\epsfbox{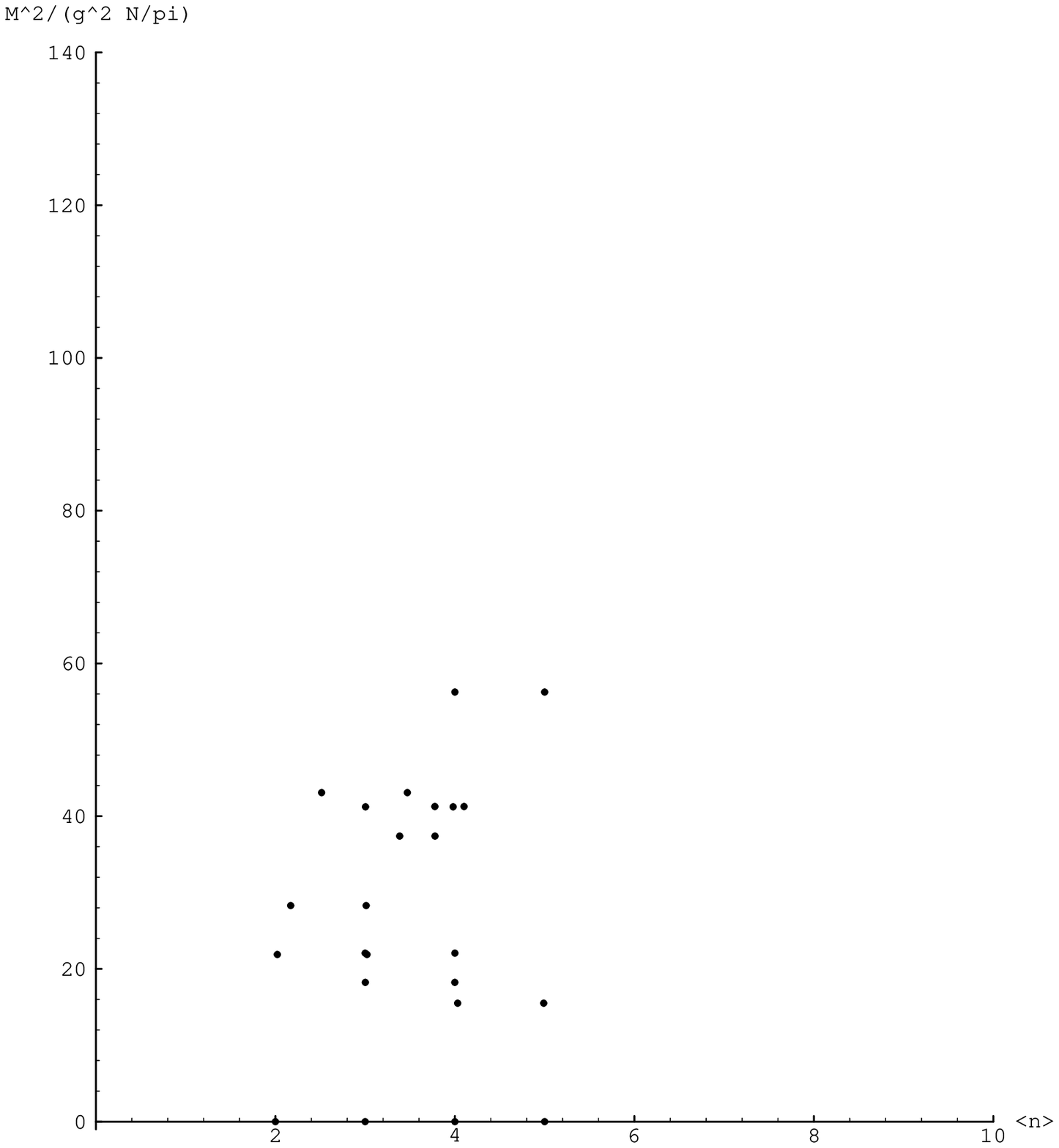}}
 \caption{
Mass squared of bosonic bound states for $K=5$ as a function of
the average number of constituents; $M^2$ are measured in units of
$g^2 N / \pi$
}
 \label{k5fig}
\end{figure}
\begin{figure}
 \leavevmode
 \epsfysize=18cm
 \centerline{\epsfbox{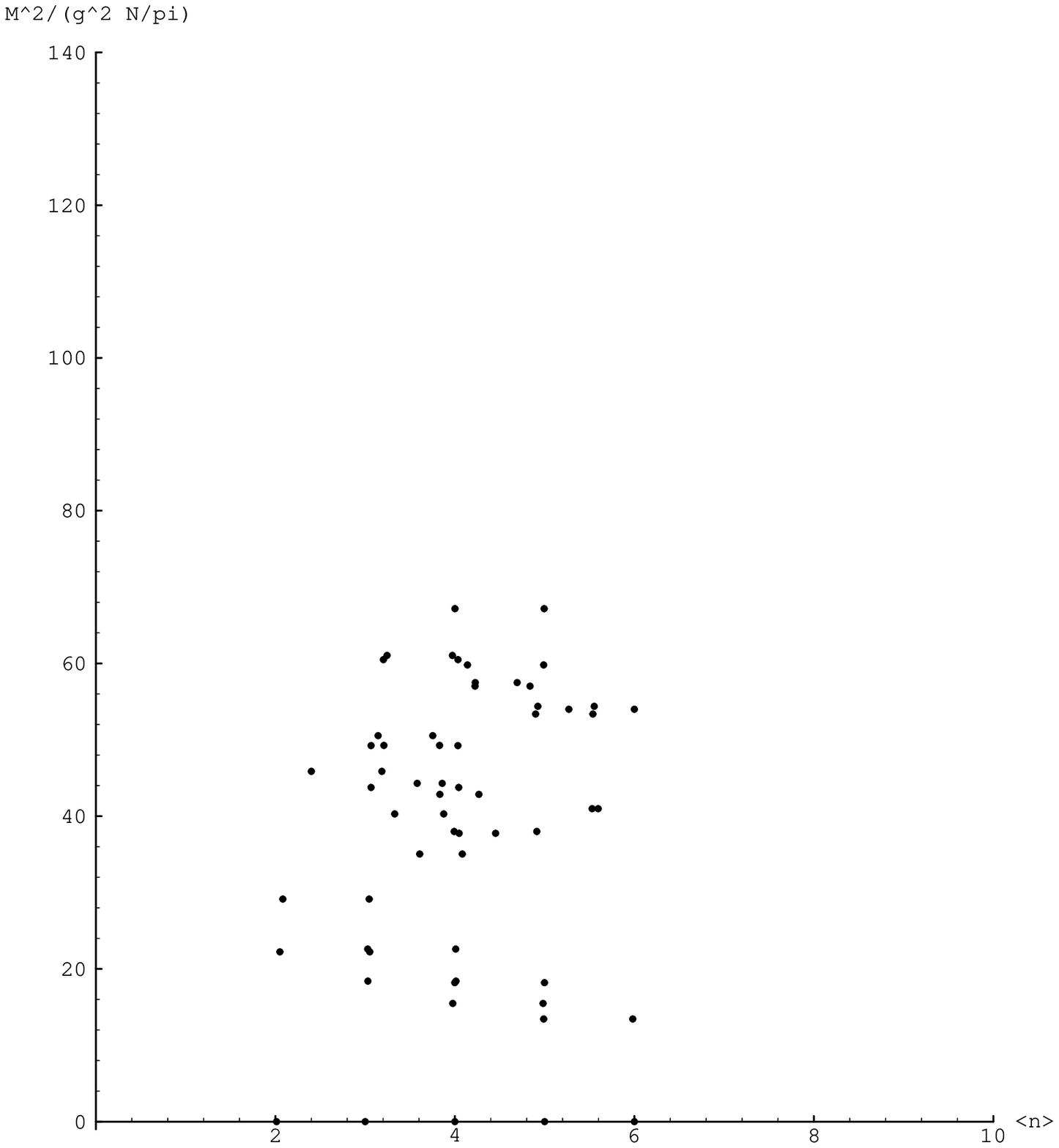}}
 \caption{
Mass squared of $K=6$ bosonic bound states as a function of
the average number of constituents; $M^2$ are measured in units of
$g^2 N / \pi$
}
 \label{k6fig}
\end{figure}

\begin{figure}
 \leavevmode
 \epsfysize=18cm
 \centerline{\epsfbox{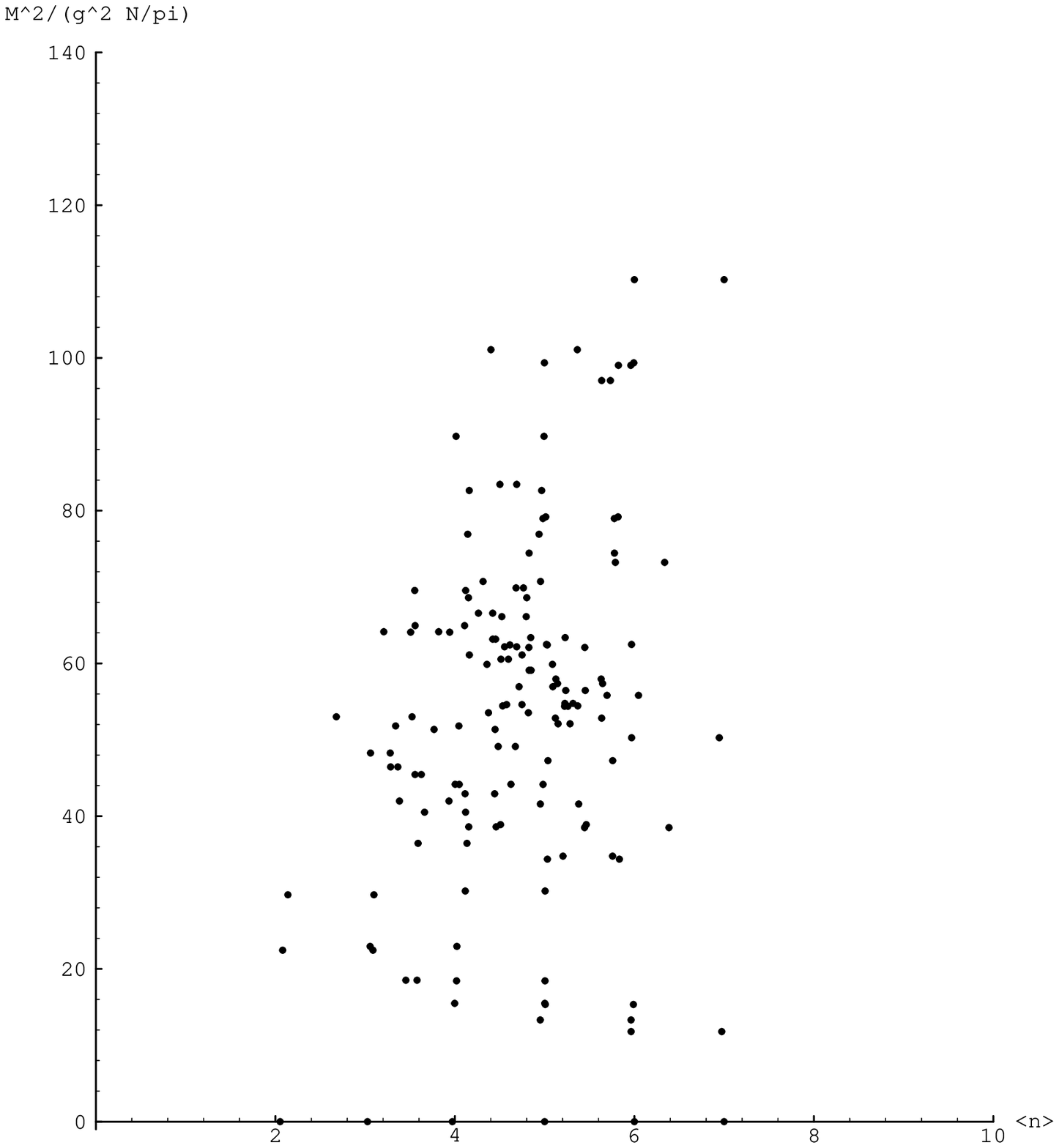}}
 \caption{
Mass squared of $K=7$ bosonic bound states as a function of
the average number of constituents; $M^2$ are measured in units of
$g^2 N / \pi$
}
 \label{k7fig}
\end{figure}

\begin{figure}
 \leavevmode
 \epsfysize=18cm
 \centerline{\epsfbox{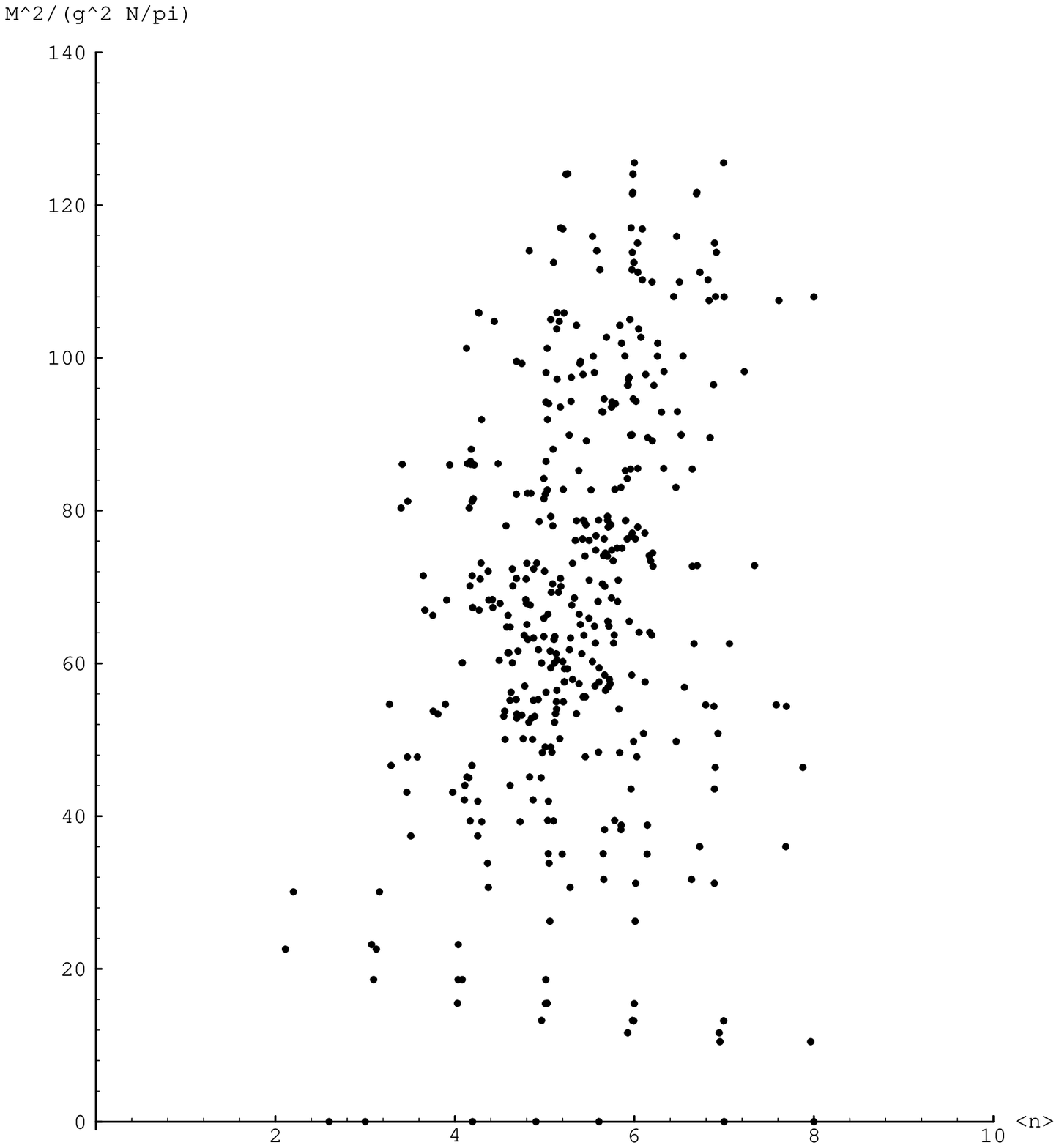}}
 \caption{
Mass squared of bosonic bound states for $K=8$ as a function of
the average number of constituents; $M^2$ are measured in units of
${g^2 N \over \pi}$.
}
 \label{k8bfig}
\end{figure}

\begin{figure}
 \leavevmode
 \epsfysize=18cm
 \centerline{\epsfbox{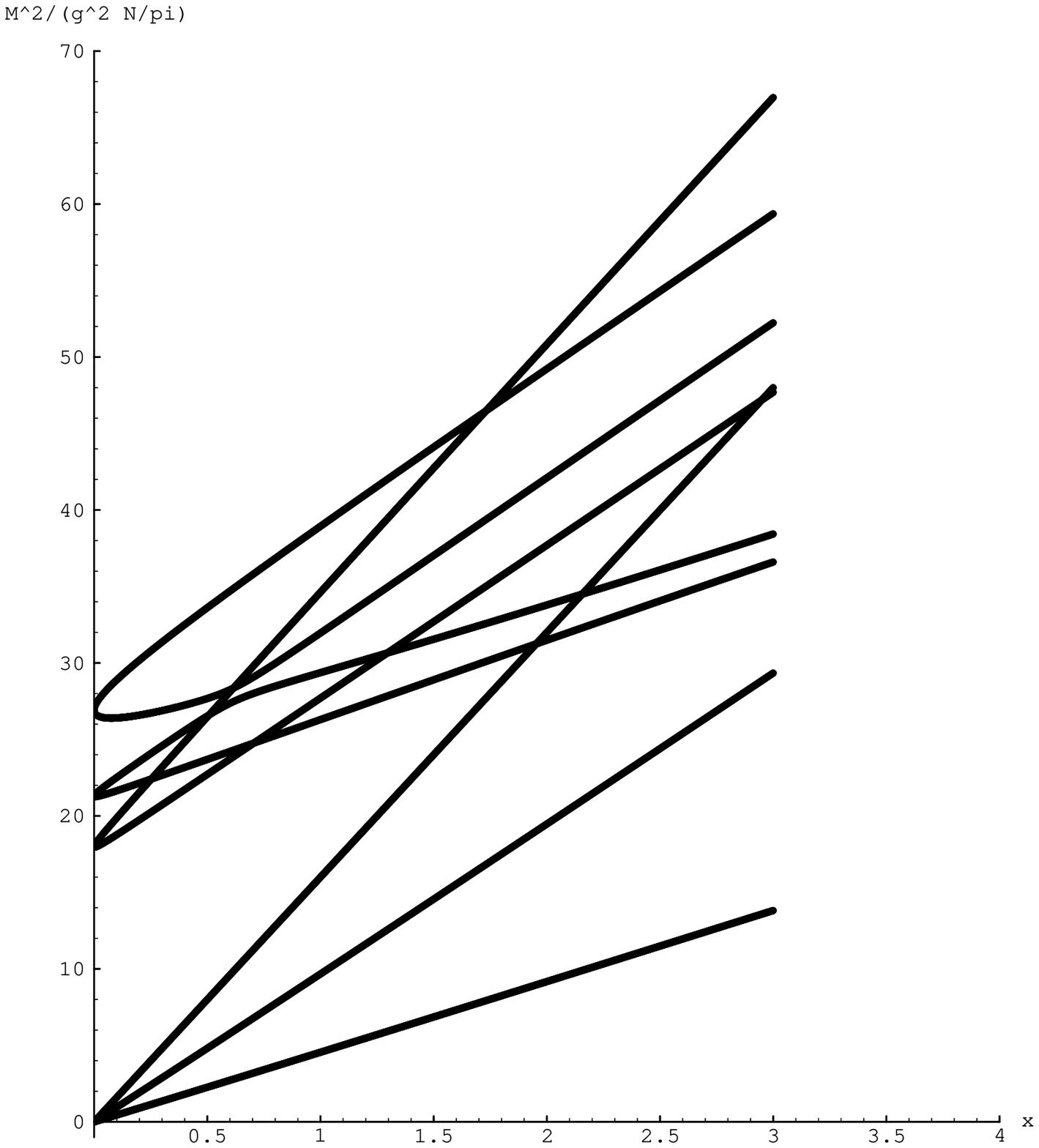}}
 \caption{
Mass squared of $K=4$ bosonic bound states as a function of
the constituent mass squared; both are measured in units of
$g^2 N / \pi$.
}
 \label{k4bmm}
\end{figure}

\begin{figure}
 \leavevmode
 \epsfysize=18cm
 \centerline{\epsfbox{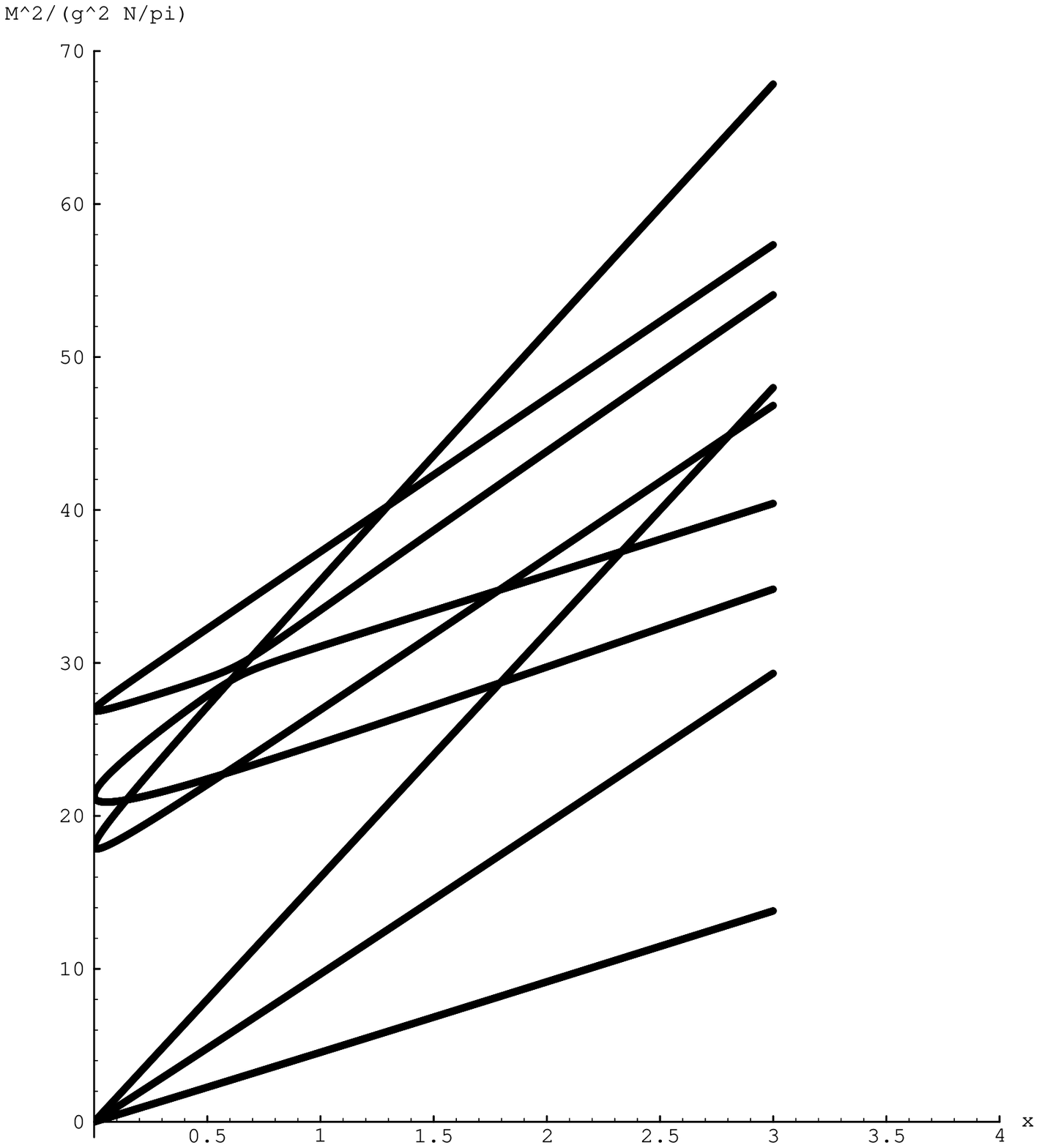}}
 \caption{
Mass squared of $K=4$ fermionic bound states as a function of
the constituent mass squared; both are measured in units of
$g^2 N / \pi$.
}
 \label{k4fmm}
\end{figure}

\end{document}